\documentclass{elsarticle}
\usepackage[utf8]{inputenc}

\title{Particle-level Simulation of Magnetorheological Fluids: A Fully-Resolved Solver}

\author[mythirdaddress,mysecondaddress,myprimaryaddress]{C. Fernandes \corref{mycorrespondingauthor}}

\author[mythirdaddress]{Salah A. Faroughi \corref{mycorrespondingauthor}}

\cortext[mycorrespondingauthor]{Corresponding authors: cbpf@dep.uminho.pt \& salah.faroughi@txstate.edu}

\address[mythirdaddress]{Geo-Intelligence Laboratory, Ingram School of Engineering, Texas State University, San Marcos, Texas, 78666, USA}
\address[mysecondaddress]{Associate Laboratory of Intelligent Systems (LASI), Institute for Polymers and Composites (IPC), Polymer Engineering Department, School of Engineering, University of Minho, Campus of Azur\'em, 4800-058 Guimar\~aes, Portugal}
\address[myprimaryaddress]{Center of Mathematics (CMAT), University of Minho, Campus of Gualtar, 4710-057 Braga, Portugal}

\date{\today}

\usepackage{mwe}

\usepackage{setspace} 
\usepackage{lineno,hyperref}
\modulolinenumbers[5]

\usepackage[a4paper]{geometry}
\usepackage{amssymb}
\usepackage[]{amsmath}
\usepackage{upgreek}

\usepackage{pgfplots}
\pgfplotsset{compat=1.5}
\usepgfplotslibrary{units}
\usetikzlibrary{spy}
\usetikzlibrary{pgfplots.groupplots}
\usepackage{bm}
\usepackage{soul,color}
\soulregister\cite7
\soulregister\ref7
\soulregister\pageref7

\usepackage{epstopdf}
\usepackage{units}
\usepackage{lscape}
\usepackage{booktabs}
\usepackage{gensymb}
\usepackage{multirow}

\usepackage{graphicx}
\usepackage{caption}
\usepackage{subcaption}

\usepackage{floatrow} 
\usepackage[flushleft]{threeparttable}
\usepackage{tikz}
\usepackage{placeins}
\usepackage{color}
\usepackage{hyperref}

\usepackage{natbib}

\usepackage{stackengine}

\usepackage{array}
\usepackage{tabularx}

\usepackage{algorithm}
\usepackage{algpseudocode}

\def\mybibdoicolor{\color{black}}
\newcommand*{\doi}[1]{\href{\detokenize{#1}} {\raggedright\mybibdoicolor{DOI: \detokenize{#1}}}}

\begin{document}

\begin{abstract}
Magnetorheological fluids (MRFs) are smart materials consisting of micro-scale magnetizable particles suspended in a carrier fluid. The rheological properties of a MRF can be changed from a fluid-state to a solid-state upon the application of an external magnetic field. This study reports the development of a particle-level simulation code for magnetic solid spheres moving through an incompressible Newtonian carrier fluid. The numerical algorithm is implemented within an open-source finite-volume solver coupled with an immersed boundary method (FVM-IBM) to perform fully-resolved simulations. The particulate phase of the MRF is modeled using the discrete element method (DEM). The resultant force acting on the particles due to the external magnetic field (i.e., magnetostatic polarization force) is computed based on the Clausius-Mossotti relationship. The fixed and mutual dipole magnetic models are then used to account for the magnetic (MAG) interactions between particles. Several benchmark flows were simulated using the newly-developed FVM-IBM-DEM-MAG algorithm to assess the accuracy and robustness of the calculations. First, the sedimentation of two spheres in a rectangular duct containing a Newtonian fluid is computed without the presence of an external magnetic field, mimicking the so-called drafting-kissing-tumbling (DKT) phenomenon. The numerical results obtained for the DKT case study are verified against published data from the scientific literature. Second, we activate both the magnetostatic polarization and the dipole-dipole forces and resultant torques between the spheres for the DKT case study. Next, we study the robustness of the FVM-IBM-DEM-MAG solver by computing multi-particle chaining (i.e., particle assembly) in a two-dimensional (2D) domain for area volume fractions of 20$\%$ (260 particles) and 30$\%$ (390 particles) under vertical and horizontal magnetic fields. Finally, the fourth computational experiment describes the multi-particle chaining in a three-dimensional (3D) domain allowing to study fully-resolved MRF simulations of 580 magnetic particles under vertical and horizontal magnetic fields. 

\end{abstract}


\maketitle

\section{Introduction}
\label{sec:Intro}

Magnetic particle suspensions, also known as magnetorheological fluids (MRFs), appear in a variety of applications \cite{Akira2017, Kumar2019}. In the traditional fluid engineering field, the magnetorheological effect has been applied to develop mechanical actuators and dampers \cite{Bullough1996, Wereley2013}. In the newly emerged bio-engineering and drug delivery fields, there have been strong efforts to synthesize magnetic-based multifunctional particles \cite{Kuz1999, Wein2013}. In addition, in the field of natural resource  and environmental engineering \cite{Girg2010, Lan2013}, precious metals or harmful substances dissolved in sea water are captured and recovered using magnetic particles subjected to an externally applied magnetic field.

When a magnetically polarizable particle is subjected to an externally applied magnetic field, they acquire dipole moments and become magnetized \cite{Chow2006, Grant2008}.  A magnetized particle starts interacting with neighboring magnetized particles leading to the formation of chain-like structures or clusters of particles aligned with the magnetic field direction (i.e., particle assembly) \cite{Kang2008}. To date, numerous studies have investigated the dynamics of MRFs under magnetic fields.
\citet{Hayes2001} studied magnetic particles in microchannels by describing reversible self-assembled regularly spaced structures, when  particles were exposed to an external magnetic field. From their study, they concluded that magnetic particles can be used in an extensive variety of on-chip applications and unique microfabrication techniques, automating the laboratory procedures. \citet{Melle2003} also developed a chain model for magnetorheological fluids in rotating magnetic fields. Through single-chain simulations as well as through experimental measurements, they showed that the chain shape and orientation depends strongly on the magnetic permeability of the particles $\mu_p$. Subsequently, \citet{Keaveny2008} developed a finite-dipole model, where the magnetization of a particle is represented as a distribution of current density. This was proposed to estimate the magnetic forces between magnetic particles accurately and efficiently such that it can be applicable for systems with thousands of particles. In their model, the induced magnetization of a particle is represented as a localized Gaussian distribution of current that is added as a source term in the Poisson equation for the vector potential of the magnetic field \cite{Kang2008}. The procedure yields very accurate solutions to collinear three-body problems. However, the scheme is not as accurate when considering other configurations with a large number of particles, because there is the need to include more information from the far field (e.g., quadrupole moments). \citet{Han2010} presented a two-stage computational procedure for the numerical modelling of magnetorheological fluids. At the first stage, the particle dynamics is modelled using the discrete element method (DEM), whereas the hydrodynamic forces on the particles are approximated simply using the Stokes' law (i.e., the fluid flow was not explicitly resolved) \cite{Stokes1851}. At the second stage, they deployed a combined approach using lattice Boltzmann method (LBM) and DEM to fully resolve the fluid fields, particle-particle, and particle–fluid hydrodynamic interactions. However, they raised an issue related to the accuracy of the magnetic interaction models while retaining the computational simplicity and efficiency. Subsequently, \citet{Ke2017} developed a fully-resolved scheme based on lattice Boltzmann, immersed boundary, and discrete element methods (LBM-IBM-DEM) to simulate the behavior of magnetic particles moving in a fluid subject to an external magnetic field. The numerical results obtained showed that the LBM-IBM-DEM scheme was able to capture the major physical features of magnetic particle's motion in a fluid. Specifically, they showed that particles first form fragmented chains along the magnetic direction. These chain-like clusters then continue to grow and align, and eventually, they approach an \textit{near} steady state configuration. Additionally, it was shown that with the increase of the magnetic field a faster particle motion or merging between short chains occurs. Recently, \citet{Zhang2019} developed a two-phase numerical simulation method using LBM-IBM-DEM approach  to investigate the yielding phenomena during the start-up process of a MRF flowing through a microchannel under a transverse uniform magnetic field. The yielding of the MRF flowing through the microchannel was studied as a proxy to the deformation of the chains composed of magnetic particles. They showed that the yielding of a single-chain at different inlet velocities was regular. However, for a multi-chain system where chains are entangled, the yielding behavior presented an unpredictable regularity. \citet{Zhou2019} also studied the motion of magnetic particles in a 3D microchannel flow modulated by the alternating gradient magnetic field. They used the LBM-IBM numerical simulation scheme, and showed that magnetic particles initially agglomerate due to their magnetic dipole force and then move together with the carrier fluid. They also showed that, in an alternating gradient magnetic field, magnetic particles oscillate along the flow direction, disturb the flow field, and increase the overall turbulence intensity. \citet{Leps2021} modeled the dynamics of MRFs using DEM method alone leveraging the open source LIGGGHTS \cite{GMBH2015Contact} software. The algorithm is based on the mutual-dipole model to allow for the use of a large number of magnetic particles with several close neighbors while keeping a good trade-off between model accuracy and computational cost. Using accurate particle size distributions, high heritage contact models, and an uncoupled fluid model, \citet{Leps2021} were able to match the experimentally derived  yield stress results for MRFs more closely than using mono-disperse particle size distributions. Lastly, \citet{Tajfirooz2021} presented an Eulerian–Lagrangian approach for simulating the magneto-Archimedes separation of neutrally buoyant non-magnetic spherical particles within MRFs. A four-way coupled point-particle method \cite{Zhou2010, Fernandes2018} was employed, where all relevant interactions between an external magnetic field, a magnetic fluid and immersed particles were taken into account. First, the motion of rigid spherical particles in a magnetic liquid was studied in single- and two-particle systems. It was shown that numerical results of single- and two-particle configurations were in good agreement with detailed experimental results on particle position. Subsequently, the magneto-Archimedes separation of particles with different mass densities in many-particle systems interacting with the fluid was also studied. It was concluded that history effects and inter-particle interactions significantly influence the levitation dynamics of particles and have a detrimental impact on the separation performance.

Most of the aforementioned numerical studies around MRFs  focus on the formation of magnetorheological structures using the simplified Stokes drag law and the dipole–dipole interaction model, excluding the hydrodynamic interactions between particles and higher order mutual magnetic interactions. The flow characteristics and chain formation features induced by coupled hydrodynamic and magnetic interactions are still missing in the literature. This is mainly  due to the lack of proper numerical models that can take into account both inter-particle magnetic and hydrodynamic interactions, in addition to other relevant attributes (e.g., particle type, size, etc.), in a fully coupled algorithm.

In this work, we develop a fully-resolved simulation  algorithm using a combination of the finite-volume, immersed boundary and discrete element methods to couple both hydrodynamic and magnetic interactions among magnetic particles suspended in Newtonian fluids. The newly-developed algorithm, so-called FVM-IBM-DEM-MAG solver, is able to describe flows with suspended magnetic particles immersed in a fluid subject to an external magnetic field. The magnetic force exerted on the particles is computed using the gradient of the magnetic field strength, which is obtained from the imposed external magnetic field \cite{Ke2017}. The magnetic interactions between the particles are implemented using a mutual dipole model \cite{Leps2021} allowing the magnetic fields of other particles to contribute to the magnetization and motion of the particle under consideration. The presented numerical algorithm has several advantages, specifically: (i) it is based on open-source libraries, OpenFOAM and LIGGGHTS, which allows the extension of the algorithm for other applications (e.g., simulation of viscoelastic fluids with suspended magnetic particles); and (ii) it employs a direct particle-level simulation methodology to resolve both hydrodynamic and magnetic interactions allowing accurate predictions of the flow patterns and particle assembly. We focus on simulations of spherical particles suspended in a Newtonian fluid in order to introduce the numerical algorithm and study its feasibility for extension to more complex flows, involving fluids with non-linear rheological behavior, and also particle with different shapes.

The remainder of this work is structured as follows. In Section~\ref{sec:methods}, we present the underlying physics and mathematical formulation describing the motion of magnetic particles in a Newtonian fluid. In Section \ref{methodology}, we present the particle-level numerical methodology leading to the FVM-IBM-DEM-MAG solver that couples the continuum and discrete phases in MRFs. 
In Section~\ref{sec:results}, we present four case studies with different level of complexities to test the developed algorithm, namely the motion and interaction of two magnetic spheres settling in an incompressible Newtonian fluid under external magnetic field, and the 2D and 3D flow behaviors of random arrays of magnetic spheres immersed in an incompressible Newtonian  fluid. Finally, in Section~\ref{sec:conclusion}, we summarize the main conclusions of this work.

\section{Underlying Physics}
\label{sec:methods}

The magnetorheological fluids (MRFs) considered in this study contain micro-scale magnetic particles with no-Brownian motion suspended in a non-magnetic incompressible Newtonian carrier fluid. MRFs deform and self-organize into mesoscopic structures depending on the internal (e.g., particle concentration) and external stimuli (e.g., temperature, flow, and magnetic fields). Among these stimuli, the application of magnetic fields is shown to provide instant action and contactless control of the mesoscopic physical structures, causing a reversible transition from a fluid-like to a solid-like state. When subjected to an  external magnetic field, particle assembly occurs that provides the fluid with the ability to transmit force. In that state, the effective viscosity of the fluid increases to the extent of becoming a viscoelastic solid.  The particle assembly promoted by magnetic field can be controlled, i.e., destroyed, deformed, or delayed. 
To accurately predict the particle assembly and chain formation in MRFs, the coupled interactions between the magnetic field, fluid, and particles must be resolved. The dynamics of MRFs, thus, present a multi-physics problem across different scales. In this section, we present the underlying physics governing the dynamics of MRFs made of rigid micro-scale magnetic spheres suspended in non-magnetic incompressible Newtonian carrier fluids under a static magnetic field.

\subsection{Magnetostatic fields}

Macroscopic electromagnetic phenomena are described using Maxwell's fundamental equations \cite{Akira2017,Chow2006}. In this study, we assume  the quantities of interest (e.g., magnetic field strength) do not vary with time, and there is no interaction between electric and magnetic fields. Therefore, we can decouple electrostatic and magnetostatic fields, and consider the problem of magnetostatic field with no free electric currents. The Maxwell's equations for  magnetostatic cases reduce to,
\begin{align}
\nabla\cdot\textbf{B}=0,
\label{eqn:divB}
\end{align}
\begin{align}
\nabla\times\textbf{H}=0,
\label{eqn:curlH}
\end{align}
where $\textbf{B}$ is the magnetic flux density, and $\textbf{H}$ is the magnetic field strength. Here $\nabla$ denotes the gradient operator, $\nabla\cdot$ denotes the divergence operator, and $\nabla\times$ denotes the curl tensor operations. For a linear isotropic domain (matrix) with a constant magnetic permeability, $\mu$, the constitutive equation relating the two field quantities, $\textbf{B}$ and $\textbf{H}$, reads as,
\begin{align}
\textbf{B}=\mu\textbf{H},
\label{eqn:constLaw}
\end{align}
where 
\begin{align}
\mu=\begin{cases}
    \mu_p  & \text{ in the particle domain,} \\
    \mu_f & \text{ in the fluid domain,}
\end{cases}
\label{eqn:permeability}
\end{align}
with $\mu_p$ and $\mu_f$ denoting the particles and base fluid's magnetic permeability, respectively. Notice that $\mu$ is discontinuous at fluid–particle interfaces, and, therefore, should be evaluated by following a similar interpolation of material properties as the one used in the level set method \cite{Kang2008, Osher2003}. Hereafter, consider that the total  computational domain is represented by $\Omega$~$=\Omega_s\cup\Omega_f$, where $\Omega_s$ is solid (``solid particles") domain, and $\Omega_f$ is the fluid domain. The total domain, solid and fluid boundaries are represented by $\partial\Omega$, $\partial\Omega_s$ and $\partial\Omega_f$, respectively.

To solve the first-order differential equations involving the two magnetic field quantities, Eqs.~(\ref{eqn:divB}) and (\ref{eqn:curlH}), we first  convert them into a second-order differential equation involving only one magnetic field quantity. For that purpose, Eq.~(\ref{eqn:curlH}) admits the existence of a scalar potential, $\phi$, such that,
\begin{align}
\textbf{H}=-\nabla\phi,
\label{eqn:BcurlA}
\end{align}
which can be substituted into Eq~(\ref{eqn:divB}) with the aid of Eq.~(\ref{eqn:constLaw}) to yield the following second-order differential equation,
\begin{align}
\nabla^2\left(\mu\phi\right)=0.
\label{eqn:secOrderEq}
\end{align}

\subsubsection{Magnetic forces and torques}
In order to describe the particle motion and the flow around it influenced by a magnetic field, a relationship between the applied magnetic field and the resultant force acting on the particles is needed. This force, known as magnetostatic polarization force \cite{Ke2017}, on particle $i$ can be evaluated as \cite{Ke2017},
\begin{equation}
\begin{aligned}
\textbf{F}_i^{me}=\int_{\Omega_s}\left(\mu_f \chi_e H \nabla H\right)~d\Omega_s,
\end{aligned}
\label{eq:magneticforce}
\end{equation}
where $\chi_e$ stands for the particle's magnetic susceptibility given by the Clausius-Mossotti relationship \cite{Jackson1999},
\begin{equation}
\begin{aligned}
\chi_e = \frac{3\mu_p}{3+\mu_p}.
\end{aligned}
\label{eq:clausiusMossotti}
\end{equation}
The torque generated by the magnetostatic polarization force on particle $i$ is computed as,
\begin{equation}
\begin{aligned}
\textbf{T}_i^{me}=\int_{\Omega_s}\left(\mu_f \chi_e H \times H\right)~d\Omega_s.
\end{aligned}
\label{eq:magnetictorque}
\end{equation}

Another fundamental force in MRFs is the magnetic dipoles evidenced by particles with opposite magnetic point poles \cite{gontijo2017numerical}. Therefore, in MRFs, particle motion is affected not only by an external magnetic field, but also by other nearby magnetized particles since each particle has a permanent magnetic moment, $\textbf{m}$. The dipole-dipole interactions between particles $i$ and $j$ results in dipole-dipole inter-particle magnetic force $(\textbf{F}_{ij}^{d-d})$ and torque $(\textbf{T}_{ij}^{d-d})$ that are calculated using the dipole-dipole contact model \cite{Ferraro1961,lammpsDipole} as,
\begin{equation}
\begin{aligned}
\textbf{F}_{ij}^{d-d} = \frac{3}{r^5}\left[\left(\textbf{m}_i\cdot\textbf{m}_j\right)\textbf{r}-\frac{5}{r^2}\left(\textbf{m}_i\cdot\textbf{r}\right)\left(\textbf{m}_j\cdot\textbf{r}\right)\textbf{r}+\left(\textbf{m}_j\cdot\textbf{r}\right)\textbf{m}_i+\left(\textbf{m}_i\cdot\textbf{r}\right)\textbf{m}_j\right],
\end{aligned}
\label{eq:dipdipforce}
\end{equation}
and
\begin{equation}
\begin{aligned}
\textbf{T}_{ij}^{d-d} = -\frac{1}{r^3}\left[\left(\textbf{m}_i\times\textbf{m}_j\right)-\frac{3}{r^2}\left(\textbf{m}_j\cdot\textbf{r}\right)\left(\textbf{m}_i\times\textbf{r}\right)\right],
\end{aligned}
\label{eq:dipdiptorque}
\end{equation}
where $\textbf{m}_i$ and $\textbf{m}_j$ are the magnetic moment vectors of the two particles, $\textbf{r}$ is the separation vector between the two particles, and $r$ is the magnitude of the separation vector $\textbf{r}$. For MRFs consisting of $N$ particles, a direct evaluation of the dipole-dipole interaction alone is $O(N^2)$ operations. This puts a severe computational constraint on the number of particles that can be simulated with a direct computation of the inter-particle dipole-dipole force. To compute the magnetic moment of each particle, $\textbf{m}$, the fixed dipole model \cite{Han2010} or the mutual dipole model \cite{Leps2021} can be used. In dilute MRFs (i.e., low concentration of magnetic particles), it is often acceptable to use the magnetic moment calculated from the background magnetic field (i.e., fixed dipole model). However, in concentrated MRFs (i.e., high concentration of magnetic particles), the induced magnetic fields from magnetized neighboring particles begin to have a significant effect on the particles magnetic moment vector. Therefore, for accuracy in force calculations, a more complex model (i.e., mutual dipole model) should be leveraged. 

\subsubsection{Fixed dipole model}

When the effect of the extra magnetic field generated by  neighboring magnetized particles is negligible on particles dynamics, then it is safe to assume that any particle is theoretically magnetized only by the externally applied magnetic field. Therefore, each particle is considered as a point dipole and the magnetic force between the particles are pairwise only. In this model, the magnetic moment of particle $i$ is given by \cite{Han2010},
\begin{equation}
\begin{aligned}
\textbf{m}_i=4\pi r^3\frac{\chi-1}{\chi+2}\textbf{H}_0,
\end{aligned}
\label{eq:fixedDipoleM}
\end{equation}
where $\chi=\mu_p/\mu_f$ is the relative susceptibility of the particles over the carrier fluid, and $\textbf{H}_0$ is the magnetic field strength of the externally applied uniform magnetic field. Notice that as the carrier fluid is assumed to be non-magnetic, its permeability is the same as that of a vacuum, i.e., $\mu_f=\mu_0=4\pi\times 10^{-7}$~[Tm/A], where T is Tesla, m is meter, and A is Ampere. This model is accurate when the two particles are far apart, and it loses accuracy when the separation distance of the particles decreases. The accuracy of the model also depends on the relative susceptibility, $\chi$.  It has been shown by \citet{Keaveny2008} that, at $\chi=5$, the fixed dipole model underestimates the maximum attractive force by around $35\%$, whereas overestimates the maximum repulsive force by $50\%$ or more, and the errors increase for larger $\chi$ values.

\subsubsection{Mutual dipole model}

The mutual dipole model \cite{Keaveny2008} allows for the magnetic fields of the neighboring  particles to contribute to the magnetization of the particle under consideration. A particle, thus, is subjected not only to the primary magnetization due to the external magnetic field, but also to a secondary magnetization from the other particles' magnetic fields. Considering the mutual magnetization of $N$ magnetizable particles with their centres at $\textbf{x}_i$ $(i=1,\cdots,N)$ in a uniform magnetic field with strength $\textbf{H}_0$, the magnetic moment of the particle $i$, $\textbf{m}_i$, is given by \cite{Leps2021},
\begin{equation}
\begin{aligned}
\textbf{m}_i=4\pi r^3\frac{\chi-1}{\chi+2}\left[\textbf{H}_0+\textbf{H}(\textbf{x}_i)\right]\qquad (i=1,\cdots,N),
\end{aligned}
\label{eq:mutualDipole}
\end{equation}
where $\textbf{H}(\textbf{x}_i)$ represents the total secondary magnetic field strength generated by other magnetized particles. The total secondary magnetic field strength can be expressed as \cite{Leps2021},
\begin{equation}
\begin{aligned}
\textbf{H}(\textbf{x}_i)=\displaystyle\sum_{j,j\neq i}^{N} \textbf{H}_j(\textbf{m}_j,\textbf{r}_{ij})=\displaystyle\sum_{j,j\neq i}^{N}\frac{1}{4\pi}\frac{3\hat{\textbf{r}}_{ij}(\textbf{m}_j\cdot \hat{\textbf{r}}_{ij})-\textbf{m}_j}{r^3_{ij}},
\end{aligned}
\label{eq:magneticSecField}
\end{equation}
with $\textbf{r}_{ij}=\textbf{x}_i-\textbf{x}_j$, $r_{ij}=|\textbf{r}_{ij}|$, and $\hat{\textbf{r}}_{ij}=\textbf{r}_{ij}/r_{ij}$. Once the $\textbf{m}_i$ values are computed for all particles, the inter-particle dipole-dipole force and torque between any two pairs are obtained using Eqs.~(\ref{eq:dipdipforce}) and (\ref{eq:dipdiptorque}), respectively.

\subsection{Incompressible  fluid flow}

In the MRFs considered in this study, the carrier fluid is considered to be a non-magnetic incompressible Newtonian fluid. The  governing equations for the flow of these fluids consist of the continuity equation,
\begin{align}
\nabla\cdot\textbf{u} = 0\quad\text{in}\quad\Omega_f,
\label{eqn:fluidcont}
\end{align}
and the Cauchy momentum equation,
\begin{align}
\rho_f\left(\frac{\partial}{\partial t}+\textbf{u}\cdot\nabla\right)\textbf{u} = -\nabla p +\eta_S\nabla^2\textbf{u}\quad\text{in}\quad\Omega_f.
\label{eqn:fluidmom}
\end{align}

Here $\rho_f$ and $\textbf{u}$ are the fluid density and velocity vector, respectively, $t$ is the time, $p$ is the pressure, and $\eta_S$ is the viscosity of the Newtonian fluid. To complete the strong mathematical form  describing the flow of MRFs, the following initial and boundary  conditions are considered,
\begin{equation}\label{BC}
\begin{cases}
    \textbf{u}(\mathbf{x},t=0) = \textbf{u}_0(\mathbf{x})  \quad\text{in}\quad\Omega_f, \\
    \textbf{u}(\mathbf{x},t) = \textbf{u}_{\partial\Omega}\quad\text{on}\quad\partial\Omega_f, \\
    \textbf{u}(\mathbf{x},t) = \textbf{u}_i \quad\text{on}\quad\partial\Omega_s,\\
    \left(-p\textbf{I}+\eta_S \left(\nabla \textbf{u} + \nabla \textbf{u}^T\right)\right)\cdot\hat{\textbf{n}}=
\textbf{\bm{$\sigma$}}_{\partial\Omega_s}\quad\text{on}\quad\partial\Omega_s.
\end{cases}
\end{equation}

In Eq.~(\ref{BC}), $\hat{\textbf{n}}$ is the outward normal unit vector to $\partial\Omega_s$, $\textbf{\bm{$\sigma$}}_{\partial\Omega_s}$ is the stress vector acting from the fluid on the solid body surface, and $\textbf{u}_i$ is the (unknown) velocity of the solid-fluid interface. The initial velocity $\textbf{u}_0$ is required to satisfy Eq.~(\ref{eqn:fluidcont}), and the boundary velocity  $\textbf{u}_{\partial\Omega}$ should satisfy the compatibility condition (last equation in Eq.~\ref{BC}) at all times.

The motion of magnetic particles is strongly affected by short-range and long-range hydrodynamic forces (drag, lift, etc.), and the resultant torques, when they are dispersed in a viscous incompressible fluid. The hydrodynamic force acting on the surface of particle $i$ can be obtained using \cite{Glowinsky2000, Fernandes2019},
\begin{equation}
\begin{aligned}
\textbf{F}_i^h=\int_{\Omega_s}\left(-\nabla p+\eta_S\nabla^2\textbf{u}\right)~d\Omega_s.\end{aligned}
\label{eq:dragforceT}
\end{equation}
The resultant hydrodynamic torques on particle $i$, denoted by $\textbf{T}_i^h$, can be then calculated by taking the cross product between the position vector $\textbf{r}$ (pointing from the fluid cell centroid to the particle centroid) and the total force from Eq.~(\ref{eq:dragforceT}) that reads as,
\begin{equation}
\begin{aligned}
\textbf{T}_i^h=\int_{\Omega_s}\Big\{\textbf{r}\times\left(-\nabla p+\eta_S\nabla^2\textbf{u}\right)\Big\}~d\Omega_s. 
\end{aligned}
\label{eq:momentforceT}
\end{equation}

The force contribution arising from pressure does not give rise to any torque contribution, due to symmetry of spherical magnetic particles. Thus, normal forces acting perpendicular to the particle surface, such as pressure, do not induce any torque. This is not the case if particle shape departs from the spherical shape (e.g., spheroids). In MRFs, particles also experience the buoyancy force, denoted by $\textbf{F}_i^g$, which is given by the weight of the displaced fluid.  The buoyancy force can be calculated as,
\begin{equation}
\begin{aligned}
\textbf{F}_i^g=\int_{\Omega_s}
(\rho_f\textbf{g})~d\Omega_s, \end{aligned}
\label{eq:gravitationalforceorg}
\end{equation}
where $\textbf{g}$ is the gravitational acceleration vector.

\subsection{Particle transient motion}

The transient motion of dispersed magnetic particles (i.e., solid phase), can be modeled using the Newton's second law of motion as,
\begin{equation}
\begin{aligned}
m_i \frac{d\textbf{U}_i^p}{dt} = \sum_{j=1}^{n_i^c}{\textbf{F}_{ij}^c}+\sum_{j=1}^{n_i^{cut}}{\textbf{F}_{ij}^{d-d}}+\textbf{F}_i^{me}+\textbf{F}_i^h+\textbf{F}_i^g,
\end{aligned}
\label{eq:particle_velocity}
\end{equation}
and
\begin{equation}
\begin{aligned}
I_i \frac{d\boldsymbol\omega^p_i}{dt} = \sum_{j=1}^{n_i^c}{\textbf{T}_{ij}^c}+\sum_{j=1}^{n_i^{cut}}{\textbf{T}_{ij}^{d-d}}+\textbf{T}^{me}_{i}+\textbf{T}_{i}^{h},
\end{aligned}
\label{eq:particle_momentum}
\end{equation}
for the conservation of linear and angular momentum of the particle $i$ with mass $m_i$ and moment of inertia $I_i$, respectively. Here, $\textbf{U}_i^p$ and $\boldsymbol\omega^p_i$ denote the translational and angular velocities of particle $i$, respectively, $\textbf{F}_{ij}^c$ and $\textbf{T}_{ij}^c$ are the contact force and contact torque resulting from the particle-particle and particle-wall interactions (with the number of total contacts, $n_i^c$, for particle $i$) that can be calculated using different contact models \cite{Renzo2004,Kloss2012}, $\textbf{F}_{ij}^{d-d}$ and $\textbf{T}_{ij}^{d-d}$ are the dipole-dipole inter-particle magnetic force and torque for a number of $n_i^{cut}$ possible interactions in the admissible cut-off region, respectively, $\textbf{F}_i^{me}$ and $\textbf{T}^{me}_{i}$ are the magnetostatic polarization force and torque due to the external magnetic field, respectively, $\textbf{F}_i^h$ and $\textbf{T}_{i}^{h}$ are the hydrodynamic force and torque acting on particle $i$, respectively, and $\textbf{F}_i^g$ is the buoyancy force.


We leverage DEM, developed by \citet{Cundall197947} and implemented in LIGGGHTS open-source library \cite{GMBH2015Contact}, to model the transient motion of dispersed magnetic particles described by Eqs.~(\ref{eq:particle_velocity}) and (\ref{eq:particle_momentum}). In DEM, multiple search algorithms are employed to identify contacting pairs of discrete particles \cite{nezami2004fast}, and different contact models are developed to integrate various mechanisms and effects such as elasticity, plasticity, viscoelasticity, friction, cohesion, damage, fracture, etc. in the contact points \cite{GMBH2015Contact}. In this study, we adopted the spring-dashpot contact model that can be extended to other non-linear models depending on the chosen stiffness and damping parameters as function of the particle overlap displacement \cite{Kloss2012}. In this model, the total contact force between particle $i$ and particle $j$ is calculated using \cite{lu2021machine}, 
\begin{equation}\label{eq3:tcontactforce}
    \mathbf{F}_{ij}^c = (\mathbf{F}_{ij}^c)_{n} + (\mathbf{F}_{ij}^c)_{t},
\end{equation}
where $(\mathbf{F}_{ij}^c)_{n}$ is the normal contact force,
\begin{equation}\label{eq4:fn}
    (\mathbf{F}_{ij}^c)_{n}  = -k_{n}\,\delta_{n}\,{\mathbf{n}} - \gamma_n\,(\mathbf{U}_{ij}^p)_{n},
\end{equation}
and $(\mathbf{F}_{ij}^c)_{t}$ is the tangential contact force,
\begin{equation}\label{eq5:ft}
    (\mathbf{F}_{ij}^c)_{t} = \textrm{min}\left(-k_{t}\,\delta_{t} - \gamma_t\,(\mathbf{U}_{ij}^p)_{t}, \,\beta_s\,| (\mathbf{F}_{ij}^c)_{n}|\;\frac{\delta_{t}}{|\delta_{t}|}\right), 
\end{equation}
with 
\begin{equation}\label{eq6:delta}
    {\delta_{t}}^{(n)} = {\delta_{t}}^{(n-1)} + (\mathbf{U}_{ij}^p)_{t}\,\Delta{t}.
\end{equation}

In Eqs.~(\ref{eq4:fn}), (\ref{eq5:ft}) and (\ref{eq6:delta}), $\mathbf{n}$ is the unit vector in the normal direction, $k_{n}$ and $k_{t}$ are the elastic stiffness for normal and tangential contacts, respectively, $\gamma_n$  and $\gamma_t$ denote the  damping coefficients in normal and tangential directions, respectively, $\delta_n$  is normal overlap displacement between two particles, $(\mathbf{U}_{ij}^p)_{n}$ and $(\mathbf{U}_{ij}^p)_{t}$ are relative velocities in normal and tangential directions of particle $i$ relative to particle $j$, respectively, with the relative velocity defined as $\mathbf{U}_{ij}^p=\mathbf{U}_{i}^p-\mathbf{U}_{j}^p$, $\beta_s$ is the sliding friction coefficient, $\delta_{t}^{(n)}$ and $\delta_{t}^{(n-1)}$ are the tangential overlap at the current and previous step, and $\Delta{t}$ is the time step.
The resultant contact torque on particle $i$ due to its contact with particle $j$, denoted by ${\textbf{T}_{ij}^c}$, can be then calculated by taking the cross product between the total contact force from Eq.~(\ref{eq3:tcontactforce}) and the position vector leading to,
\begin{equation}\label{eq7}
    {\textbf{T}_{ij}^c} = \textbf{F}_{ij}^c\;\times\;(\mathbf{x}_c - \mathbf{x}_i),
\end{equation}
where $\mathbf{x}_c$ and $\mathbf{x}_i$ are the position of contact point and particle $i$ centroid, respectively.

\section{Numerical Methodology}\label{methodology}

This section presents the numerical formulation for an algorithm using the FVM, IBM and DEM that is able to efficiently handle the rigid body motion of magnetic spherical particles surrounded by a Newtonian fluid. The algorithm considers a fictitious domain formulation, which provides a rigorous basis for the immersed boundary (IB) implementation performed in the open source framework code $CFDEMcoupling$ \cite{Fernandes2019,Hager2014,Kenneth2017}. The open source IB solver originally developed by \citet{Hager2014} is modified and improved for this study to take into account
both hydrodynamic and magnetic interactions between the fluid continuum phase and the particulate disperse phase in a fully coupled manner. Algorithm~\ref{alg1}
 summarizes the so-called FVM-IBM-DEM-MAG solver describing the solution procedure of the fluid phase and magnetic field equations, the DEM approach to handle the particle's motion, and the IBM scheme to fully couple the continuum phase with the particulate phase.
 
\begin{algorithm}
\caption{Fully-resolved FVM-IBM-DEM-MAG algorithm to model magnetorheological fluids}
\label{alg1}
\raggedright \textbf{step 1:} at time $t=0$
\begin{itemize}
\item[(a)] Set initial and boundary conditions
\item[(b)] Send initial particle position and velocities to CFD solver from DEM solver
\end{itemize}
\textbf{step 2:} at time $t=t+\Delta t$
\begin{itemize}
\item[(a)] Compute particle volume fraction
\item[(b)] Dynamic mesh refinement
\item[(c)] Calculate loads on particles (hydrodynamic and magnetic external forces, $\textbf{F}_i^{h}$ and $\textbf{F}_i^{me}$, respectively, and torques, $\textbf{T}_i^{h}$ and $\textbf{T}_i^{me}$, respectively; particle-particle contact force and torque, $\textbf{F}_{ij}^{c}$ and $\textbf{T}_{ij}^{c}$, respectively; magneto dipole-dipole force and torque, $\textbf{F}_{ij}^{d-d}$ and $\textbf{T}_{ij}^{d-d}$, respectively; and buoyancy force $\textbf{F}_i^g$, etc.) given by

$\textbf{F}_i^h=\sum_{c\in \overline{T}_h}(-\nabla p + \eta_S\nabla^2\textbf{u})(c)\cdot V(c)\qquad\qquad\qquad\hspace{0.65cm}\textbf{F}_i^{me}=\sum_{c\in \overline{T}_h}(\mu_0\chi_e\textbf{H}\nabla \textbf{H})(c)\cdot V(c)$

$\textbf{T}_i^h=\sum_{c\in \overline{T}_h}\left[\textbf{r}(c)\times(-\nabla p + \eta_S\nabla^2\textbf{u})(c)\right]\cdot V(c)\qquad\qquad\textbf{T}_i^{me}=\sum_{c\in \overline{T}_h}(\mu_0\chi_e\textbf{H}\times\textbf{H})(c)\cdot V(c)$

$\textbf{F}_{ij}^{c}$ and $\textbf{T}_{ij}^{c}$ are calculated using the non-linear elastic Hertz-Mindlin contact model.

$\textbf{F}_{ij}^{d-d} = \frac{3}{r^5}\left[\left(\textbf{m}_i\cdot\textbf{m}_j\right)\textbf{r}-\frac{5}{r^2}\left(\textbf{m}_i\cdot\textbf{r}\right)\left(\textbf{m}_j\cdot\textbf{r}\right)\textbf{r}+\left(\textbf{m}_j\cdot\textbf{r}\right)\textbf{m}_i+\left(\textbf{m}_i\cdot\textbf{r}\right)\textbf{m}_j\right]$

$\textbf{T}_{ij}^{d-d}=-\frac{1}{r^3}\left[\left(\textbf{m}_j\times\textbf{m}_i\right)-\frac{3}{r^2}\left(\textbf{m}_i\cdot\textbf{r}\right)\left(\textbf{m}_j\times\textbf{r}\right)\right]$

$\textbf{F}_i^g=\sum_{c\in \overline{T}_h}
(\rho_f\textbf{g})(c) \cdot V(c)$

\item[(d)] Solve Newton-Euler equations (Velocity-Verlet integration) to obtain new particle position, and linear and angular velocities ($\text{in}~\Omega_p$)

$m_i \frac{d\textbf{U}_i^p}{dt} = \sum_{j=1}^{n_i^c}{\textbf{F}^{c}_{ij}}+\sum_{j=1}^{n_i^{cut}}\textbf{F}^{d-d}_{ij}+\textbf{F}^{me}_i+\textbf{F}^h_i+\textbf{F}^g_i \qquad\qquad I_i \frac{d\boldsymbol\omega_i}{dt} = \sum_{j=1}^{n_i^c}{\textbf{T}^{c}_{ij}}+\sum_{j=1}^{n_i^{cut}}\textbf{T}^{d-d}_{ij}+\textbf{T}^{me}_{i}+\textbf{T}^{h}_{i}$

\item[(e)] Solve fluid governing equations subjected to an external magnetic field ($\text{in}~\Omega_f$)

$\nabla^2\left(\mu\phi\right)=0$

$\nabla\cdot\textbf{u} = 0$

$\rho_f\left(\frac{\partial\textbf{u}}{\partial t}+\textbf{u}\cdot\nabla\textbf{u}\right)= -\nabla p +\eta_S\nabla^2\textbf{u}$

\item[(f)] Impose the rigid-body motion of the particles on the fluid velocity field
\item[(g)] Correct velocity and pressure fields

\end{itemize}

\end{algorithm}

\begin{figure}[H]
\centering
\includegraphics[width=0.7\columnwidth]{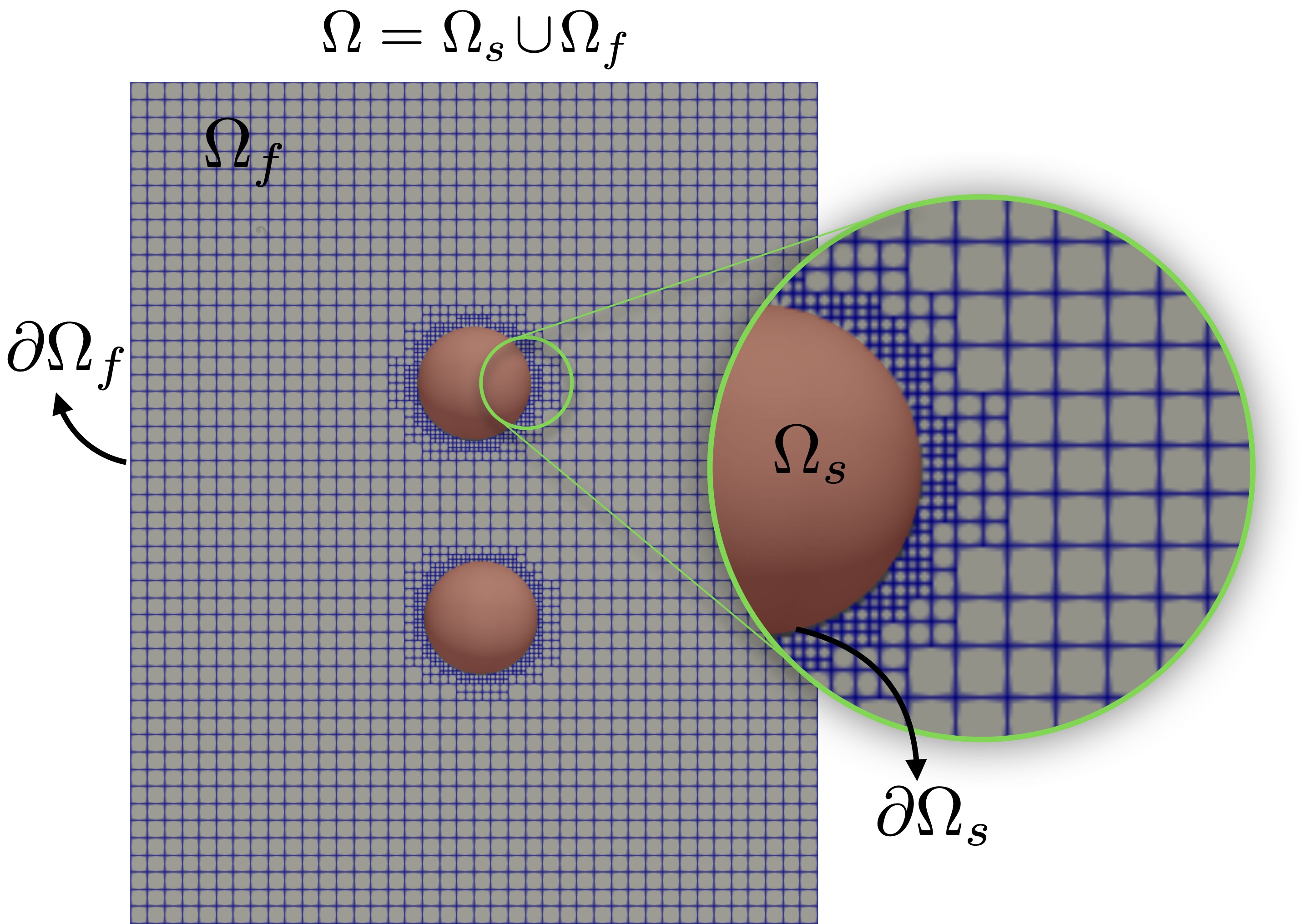}%
\caption[]
{Typical immersed boundary computational mesh configuration using dynamic refinement of the control-volumes (cells) near the particles' surface. $\Omega_f$ and $\Omega_s$ are the fluid and solid domains, respectively, with boundaries denoted by $\partial\Omega_f$ and $\partial\Omega_s$.} 
\label{fig:IBMesh}
\end{figure}

At time $t=0$, the fluid and particle initial and boundary conditions are read from the case study input files (step 1(a) in Algorithm~\ref{alg1}). Additionally, the DEM solver sends the particle initial position and velocities to the CFD solver (step 1(b) in Algorithm~\ref{alg1}). At time $t=t+\Delta t$, the numerical algorithm starts with the location of the magnetic particles, saving the cell ID of the centre position of each particle. This procedure, then, allows  to compute the particle volume fraction in each cell (step 2(a) in Algorithm~\ref{alg1}). Subsequently, as shown in Fig.~\ref{fig:IBMesh}, the algorithm uses dynamic mesh refinement near the particles' surface ($\partial\Omega_s$) to accurately capture the fluid (domain $\Omega_f$) forces developed on those regions (step 2(b) in Algorithm~\ref{alg1}).

Using the fluid solution from the last time-step in the regions marked by the particle volume fraction, the hydrodynamic, magnetostatic polarization, and buoyancy forces, $\textbf{F}_{i}^h, \textbf{T}_{i}^h, \textbf{F}_{i}^{me}, \textbf{T}_{i}^{me}, \textbf{F}_{i}^{g}$, that act on each particle's surface are computed  (step 2(c) in Algorithm~\ref{alg1}). The hydrodynamic force acting on the surface of particle $i$, denoted by $\textbf{F}_i^h$ and defined by Eq.~(\ref{eq:dragforceT}), can be rewritten as,
\begin{equation}
\begin{aligned}
\int_{\Omega_s}\left(-\nabla p+\eta_S\nabla^2\textbf{u}\right)~d\Omega_s=\int_{\Omega}\left(-\nabla p+\eta_S\nabla^2\textbf{u}\right)\delta_{\Omega}~d\Omega,\end{aligned}
\label{eq:stressinterfacesdiscretized}
\end{equation}
where, $\textbf{x}$ is an arbitrary region within the domain $\Omega$, and $\delta_\Omega = 1$ if $\textbf{x}\in\Omega_s$, otherwise $\delta_\Omega = 0$. Assuming that $T_h$ is a decomposition of $\Omega$ consisting of computational cells $c$, we can approximate Eq.~(\ref{eq:stressinterfacesdiscretized}) as, 
\begin{equation}
\begin{aligned}
\int_{\Omega}\left(-\nabla p+\eta_S\nabla^2\textbf{u}\right)\delta_{\Omega}~d\Omega=\sum_{c\in T_h}
\int_{V(c)}\left(-\nabla p+\eta_S\nabla^2\textbf{u}\right)\delta_{\Omega}~dV(c), \end{aligned}
\label{eq:stressinterfacesfinal}
\end{equation}
where $V(c)$ is the volume of cell $c$. Notice that for notation purposes we use the parentheses $(c)$ to evaluate a function on cell $c$. Numerical integration of Eq.~(\ref{eq:stressinterfacesfinal}) leads to the final form of the hydrodynamic forces acting on the particle,
\begin{equation}
\begin{aligned}
\textbf{F}_i^h=\sum_{c\in \overline{T}_h}
\left(-\nabla p+\eta_S\nabla^2\textbf{u}\right)(c)\cdot V(c),\end{aligned}
\label{eq:dragforce}
\end{equation}
where $\overline{T}_h$ is the set of all cells covered, in full or in part, by a magnetic particle. The resultant hydrodynamic torque on particle $i$, denoted by $\textbf{T}_i^h$ and defined by Eq.~(\ref{eq:momentforceT}), can be then approximated by taking the cross product between the position vector $\textbf{r}$ and the total force from Eq.~(\ref{eq:dragforce}) that reads as,
\begin{equation}
\begin{aligned}
\textbf{T}_i^h=\sum_{c\in \overline{T}_h}
\Big\{\textbf{r}(c)\times\left(-\nabla p+\eta_S\nabla^2\textbf{u}\right)(c)\Big\}\cdot V(c).\end{aligned}
\label{eq:momentforce}
\end{equation}
Similarly, the magnetostatic polarization force and torque, defined by Eqs.~(\ref{eq:magneticforce}) and (\ref{eq:magnetictorque}), are approximated numerically as, 
\begin{equation}
\begin{aligned}
\textbf{F}_i^{me}\sum_{c\in \overline{T}_h}(\mu_0\chi_e\textbf{H}\nabla \textbf{H})(c)\cdot V(c),\end{aligned}
\label{eq:magforcenum}
\end{equation}
and
\begin{equation}
\begin{aligned}
\textbf{T}_i^{me}=\sum_{c\in \overline{T}_h}(\mu_0\chi_e\textbf{H}\times\textbf{H})(c)\cdot V(c).\end{aligned}
\label{eq:magtorquenum}
\end{equation}

The buoyancy force, defined by Eq.~(\ref{eq:gravitationalforceorg}), can be also approximated numerically by integrating the fluid density $(\rho_f)$ over the volume of the solid region in the mesh, i.e., $V(c)$ with $c\in\overline{T}_h$, to obtain the total displaced fluid mass, i.e., $\rho_f V(c)$ with $c\in\overline{T}_h$. Next, by multiplying the fluid mass by the gravitational acceleration vector $(\textbf{g})$, the buoyancy force can be calculated as,
\begin{equation}
\begin{aligned}
\textbf{F}_i^g=\sum_{c\in \overline{T}_h}
(\rho_f\textbf{g})(c) \cdot V(c). \end{aligned}
\label{eq:gravitationalforce}
\end{equation}

As the next step in the FVM-IBM-DEM-MAG algorithm, the resulting forces and torques for each particle are returned to the DEM solver. Additionally, if collision between particles or particle-wall are detected, the collision force and torque, $\textbf{F}_{ij}^c$ and  $\textbf{T}_{ij}^c$, are calculated using Eqs.~(\ref{eq3:tcontactforce})--(\ref{eq7}). Finally, the dipole-dipole magnetic force and torque, $\textbf{F}_{ij}^{d-d}$ and $\textbf{T}_{ij}^{d-d}$, are calculated using Eqs.~(\ref{eq:dipdipforce}) and (\ref{eq:dipdiptorque}) with either the fixed dipole model for dilute suspensions or the mutual dipole model for non-dilute suspensions to retrieve the particle dipole moment.

A data exchange model is also used to run a DEM script, which  computes the particles' positions, translational and angular velocities (Eqs.~(\ref{eq:particle_velocity})--(\ref{eq:particle_momentum})), using Velocity-Verlet integration \cite{Verlet1967} (step 2(d) in Algorithm~\ref{alg1}). The particles' new positions and velocities are then transferred to the CFD solver. The CFD solver proceeds with the PISO (Pressure-Implicit with Splitting of Operators) algorithm \cite{Issa1986} (step 2(e) in Algorithm~\ref{alg1}), which solves the magneto-static potential equation, Eq.~(\ref{eqn:secOrderEq}), and  fluid flow governing equations, Eqs.~(\ref{eqn:fluidcont})--(\ref{BC}). An intermediate velocity field $\widehat{\textbf{u}}$ is first obtained by solving the momentum balance equations, Eq.~(\ref{eqn:fluidmom}), and then an intermediate pressure $\widetilde{p}$ is obtained from the continuity equation, Eq.~(\ref{eqn:fluidcont}), which results in a Poisson equation for the pressure correction.

The next step is to correct the intermediate velocity field $\widehat{\textbf{u}}$ in the particle region by imposing the rigid body velocity provided by the DEM calculation (step 2(f) in Algorithm~\ref{alg1}). This correction is equivalent to adding a body force per unit volume  defined as,
\begin{equation}
\begin{aligned}
\textbf{f}=\rho\frac{\partial}{\partial t}(\widetilde{\textbf{u}}-\widehat{\textbf{u}}),
\end{aligned}
\label{eq:intvelcorrected}
\end{equation}
in the momentum balance equations, Eq.~(\ref{eqn:fluidmom}), to obtain a corrected velocity field $\widetilde{\textbf{u}}$. Here  $\widetilde{\textbf{u}}=\textbf{U}_i^{p}+\boldsymbol\omega_i\times \textbf{r}$ is defined only for the cells within the solid body. The translational and angular velocities, $\textbf{U}_i^{p}$ and $\boldsymbol\omega_i$, respectively, were previously computed in step 2(d).

The previous step introduces a discontinuity in the velocity field at the interface, giving rise to a non-zero divergence in that location. Hence, the velocity field $\widetilde{\textbf{u}}$ and the pressure field $\widetilde{p}$ need to be corrected (step 2(g) in Algorithm~\ref{alg1}). For that purpose, $\widetilde{\textbf{u}}$ is projected onto a divergence-free velocity space, $\overline{\textbf{u}}$, by using a scalar field $\psi$, as:
\begin{equation}
\begin{aligned}
\overline{\textbf{u}}=\widetilde{\textbf{u}}-\nabla\psi,
\end{aligned}
\label{eq:divfreevelcorrected}
\end{equation}
where $\psi$ is obtained by solving the following Poisson equation,
\begin{equation}
\begin{aligned}
\nabla^2\psi=\nabla\cdot\widetilde{\textbf{u}}.
\end{aligned}
\label{eq:divfreepoissoneq}
\end{equation}
Then $\overline{\textbf{u}}$ is calculated by Eq.~(\ref{eq:divfreevelcorrected}). The last step is equivalent to adding a pressure force $-\rho\frac{\nabla\psi}{\Delta t}$ in the momentum conservation equations, which requires the pressure field to be corrected by,
\begin{equation}
\begin{aligned}
p=\widetilde{p}+\frac{\psi}{\Delta t}.
\end{aligned}
\label{eq:pressurecorrected}
\end{equation}

This new FVM-IBM-DEM-MAG solver is implemented within the $CFDEMcoupling$ \cite{CFDDEMcoupling2011} framework.

\section{Results and discussion}
\label{sec:results}

This section presents the validation of the proposed FVM-IBM-DEM-MAG solver against several benchmark case studies. The first case study is devoted to the sedimentation of two sphere's in a rectangular duct containing a Newtonian fluid, mimicking the so-called drafting-kissing-tumbling (DKT) phenomenon. We start by turning off the external magnetic field to verify the solver's capabilities to simulate the motion and interaction of the two settling spheres. Subsequently, in the second case study, we activate both the external magnetic field and the dipole-dipole force (and resultant torque) between the spheres for the DKT problem. This case study  allows us to test the implementation of the magnetic force acting on the particles induced by the external magnetic field and  the nearby magnetized particles. The magnetic potential gradient is applied in the vertical and horizontal directions to verify the ability of the algorithm to predict particle chaining in both directions. The third case study  tests the robustness of the FVM-IBM-DEM-MAG solver by computing multi-particle chaining with 260 and 390 spheres whose centers are located in a 2D plane. Finally, the fourth case study describes the multi-particle chaining when particles are randomly distributed in a 3D domain.

\subsection{DKT phenomenon under zero magnetic field}
\label{sec:CS1}

\begin{figure}[H]
\centering
\includegraphics[width=0.35\columnwidth]{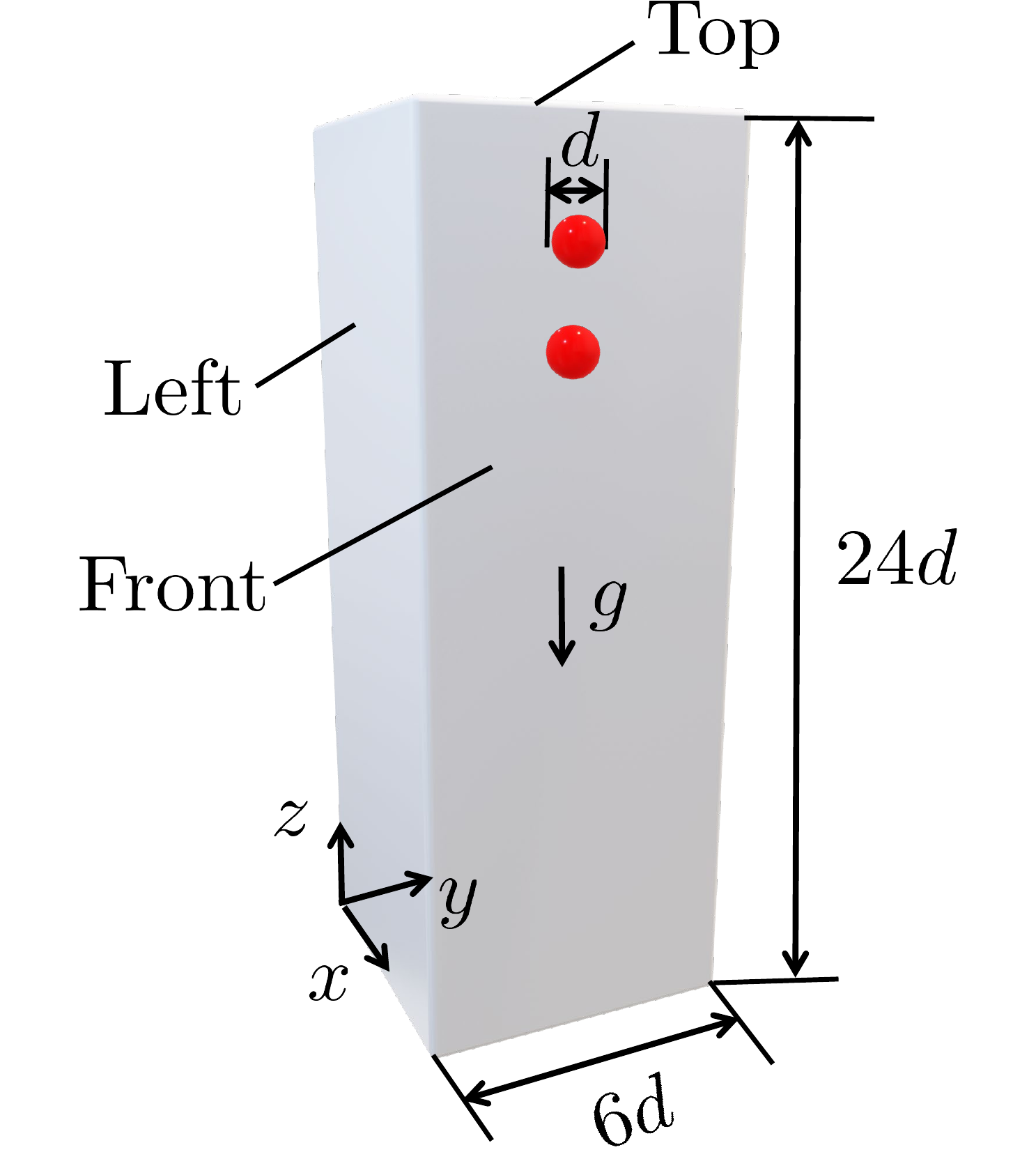}%
\caption[]
{Configuration of the drafting-kissing-tumbling (DKT) benchmark case study, where the transient motion of two spheres is considered while settling through an initially quiescent viscous fluid confined in a duct of width $6d$ and height $24d$, where $d=1/6$~cm is the sphere diameter. The schematic diagram illustrates the computational domain including the coordinate system, the boundary walls, the gravitational acceleration $g$, and the initial positions of the spheres located on $(0.5, 0.5, 3.5)$ and $(0.5, 0.49, 3.16)$.} 
\label{fig:DKTGeom}
\end{figure}

The objective of this test case is to simulate the motion and the interaction of two equal rigid spheres settling in a duct  as shown in Fig.~\ref{fig:DKTGeom}. The spherical particles are placed vertically with a distance equal to  four particle's radius. The leading sphere (i.e., the one in below) is slightly off-centred to avoid the symmetric solution. In this case, we expect the simulations to reproduce the well-documented DKT phenomenon, which has been observed in laboratory experiments \cite{Fortes1987} and modeled through numerical simulations using different computational methods \cite{Ke2017, Hu1992, Johnson1992, Feng1994}. This benchmark is specifically selected  to test the accuracy and effectiveness of the FVM-IBM-DEM-MAG algorithm, when the magnetic field is set to zero \cite{Fernandes2019, Hager2014}.

The computational domain is $\Omega = \left[0,1\right]\times\left[0,1\right]\times \left[0,4\right]$ cm$^3$. The diameter of the spheres is $d=1/6$~cm. The initial positions of the spheres centers are $(0.5, 0.5, 3.5)$ and $(0.5, 0.49, 3.16)$, and the fluid and spheres are initially at rest. On the boundary of the channel, no-slip fluid velocity is imposed. The fluid density is $\rho_f = 1$~g/cm$^3$, the sphere's density is $\rho_s = 1.14$~g/cm$^3$, and the fluid kinematic viscosity is $\nu = 0.01$~cm$^2$/s~\cite{Glowinsky2000}. For the potential inter-particle contacts and particle-wall contacts, the coefficient of normal restitution, coefficient of friction, Poisson's ratio and  Young's modulus are considered to be $0.97$, $0.10$, $0.45$, and  $2\times 10^9$ Pa, respectively. 

The numerical experiments  were performed using two hexahedral meshes with initial configuration M1: $40\times 40\times 160$ and M2: $60\times 60\times 240$ grid cells. In addition, dynamic mesh capability ($dynamicRefineFvMesh$) \cite{Jasak2009} is used to refine the mesh near the solid-fluid interface at each time-step. In this work, the maxRefinement parameter (a property of the dynamic mesh method defining the maximum number of layers of refinement that a cell can experience) is equal to two layers. The simulation time-step is set to $\Delta t = 10^{-4}$~s corresponding to an average Courant number of $0.1$. The total computational elapsed time for the simulations was 1h52m and 6h16m for M1 and M2, respectively, executed on a 3.00-GHz 48 cores Intel Xeon Gold 6248R CPU processor with 128 GB of RAM.

\begin{figure}[H]
\captionsetup[subfigure]{justification=justified,singlelinecheck=false}
    \centering
    {\renewcommand{\arraystretch}{0}
    \begin{tabular}{c@{}c}
    \begin{subfigure}[b]{.5\columnwidth}
        \centering
        \caption{{}}
        \includegraphics[width=1.0\columnwidth]{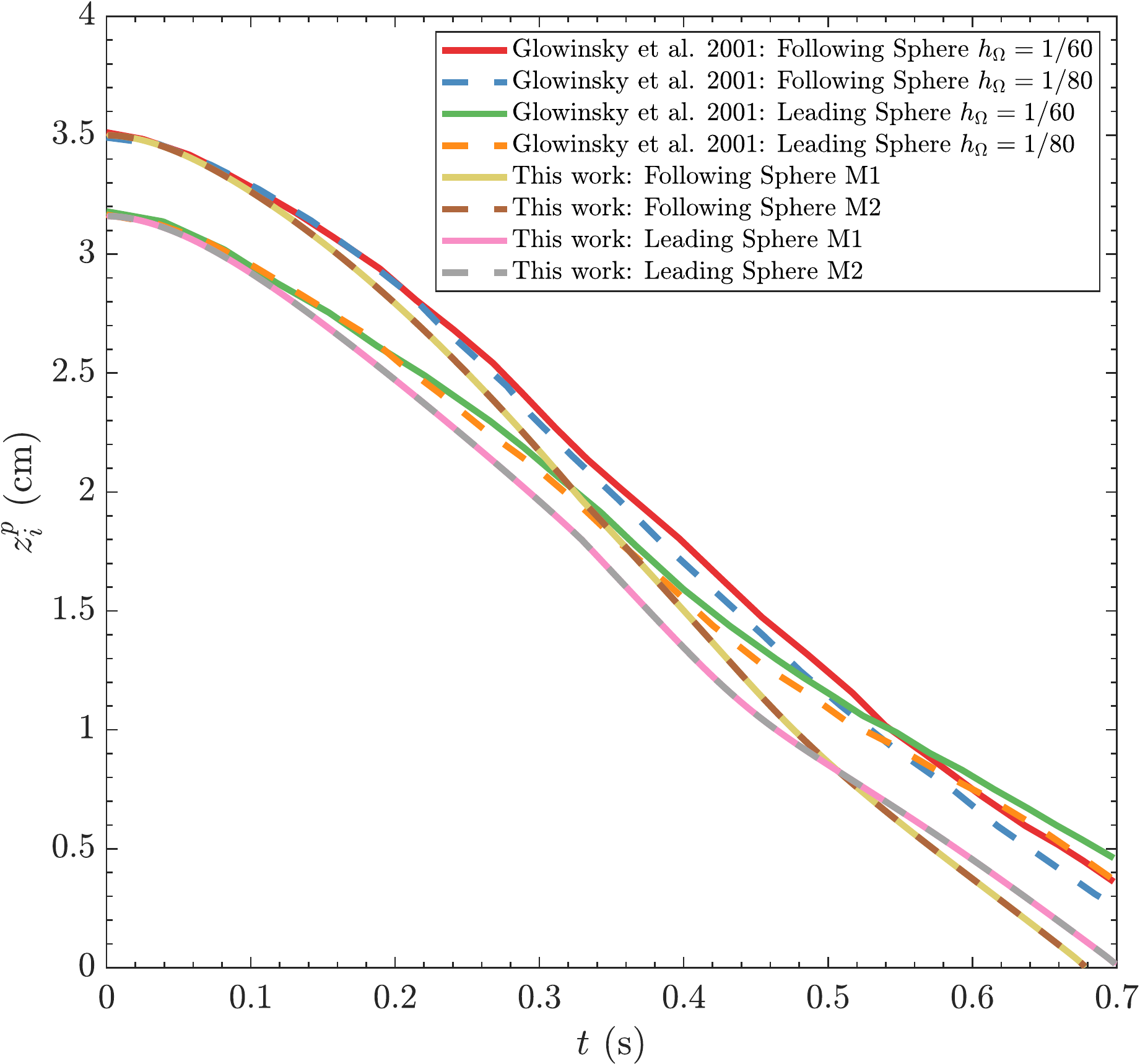}%
        \label{}
    \end{subfigure}\\
        \begin{subfigure}[b]{.5\columnwidth}
        \centering
        \caption{{}}
        \includegraphics[width=1.0\columnwidth]{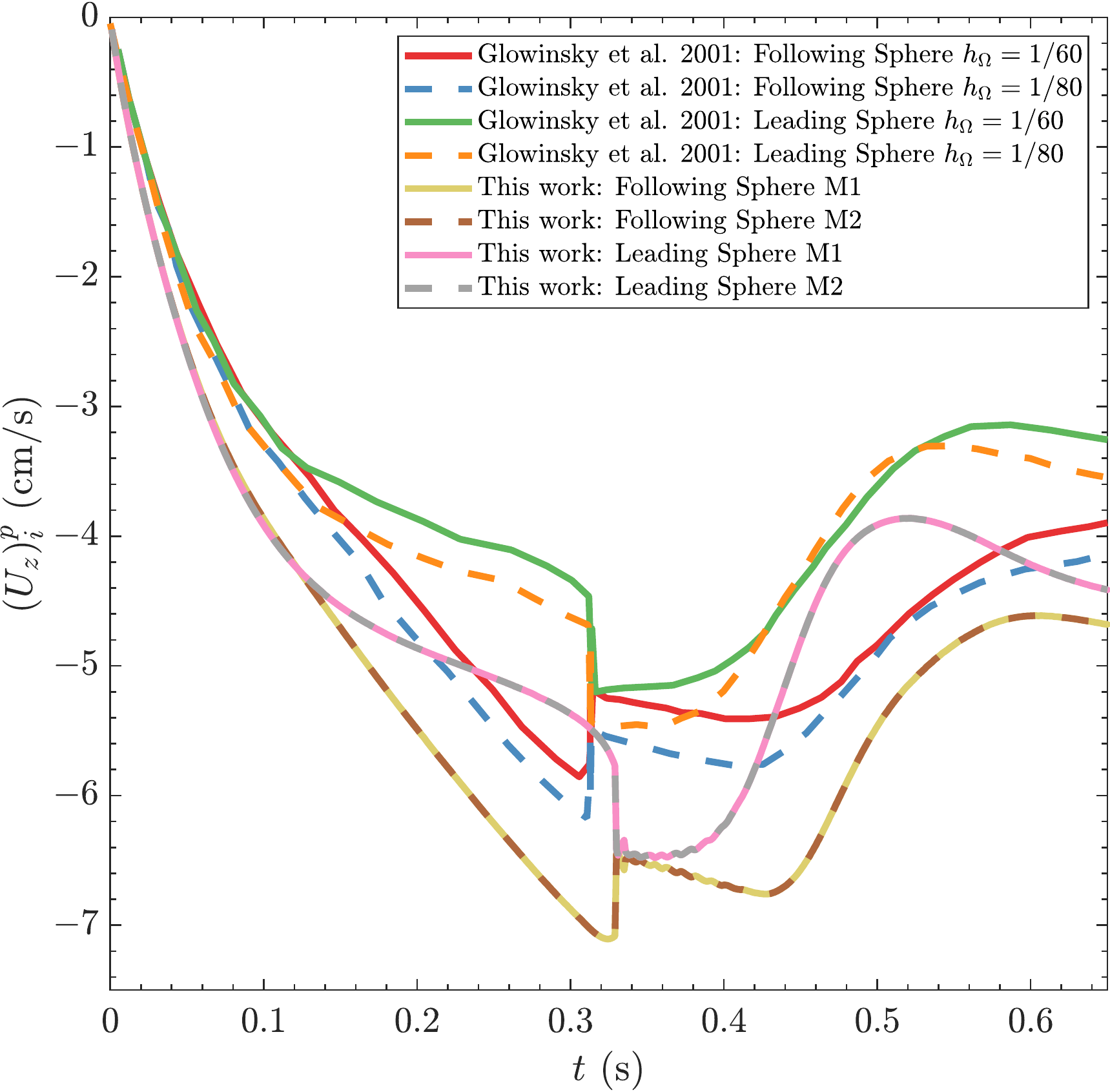}%
        \label{}
    \end{subfigure}\\
    \end{tabular}}
    \caption[]
    {A comparison of the $z$-component of the spheres' (a) center location, and (b) translation velocity as function of time for the drafting-kissing-tumbling (DKT) benchmark case study obtained using Algorithm~\ref{alg1} and those computed by \citet{Glowinsky2000}.} 
    \label{fig:DKTGlowOpenFOAM}
\end{figure}

Figure \ref{fig:DKTGlowOpenFOAM} shows the $z$-component of the spheres centers, $z_i^p$, and the $z$-component of the spheres translation velocities, $(U_z)_i^p$, as function of time for calculations using M1 and M2 meshes. Additionally, the results obtained by \citet{Glowinsky2000} using two levels of mesh refinement, $h_{\Omega}=1/60$ and $h_{\Omega}=1/80$, are included for comparison purposes. Our results for M1 and M2 meshes obtained using the newly-developed algorithm (Algorithm~\ref{alg1}) are in good agreement indicating that the results are mesh independent. As can be observed, the particle on top (following particle) is first carried by the wake generated by the particle on the bottom (leading particle) forcing to the so-called drafting phenomenon ($0\leq t < 0.14$~s). Then, the following particle velocity increases, the distance between the two particle's centres decreases, and ultimately a contact forms between them forcing to the so-called kissing phenomenon ($0.14 < t < 0.35$~s). Since the vertical configuration is unstable and particles cannot stay attached \cite{Huang1994}, the particles start tumbling and are found side by side, which is known as tumbling phenomenon ($0.35 < t < 0.5$~s). Subsequently, the following particle passes ahead of the leading particle causing the deviation of the leading particle from the middle of the channel influenced by the fluid's back-flows along the wall. Ultimately, the particle stagnate against the wall ($t\approx 0.65$~s) \cite{Ritz1999}.  

\begin{figure}[H]
\centering
\includegraphics[width=1.0\columnwidth]{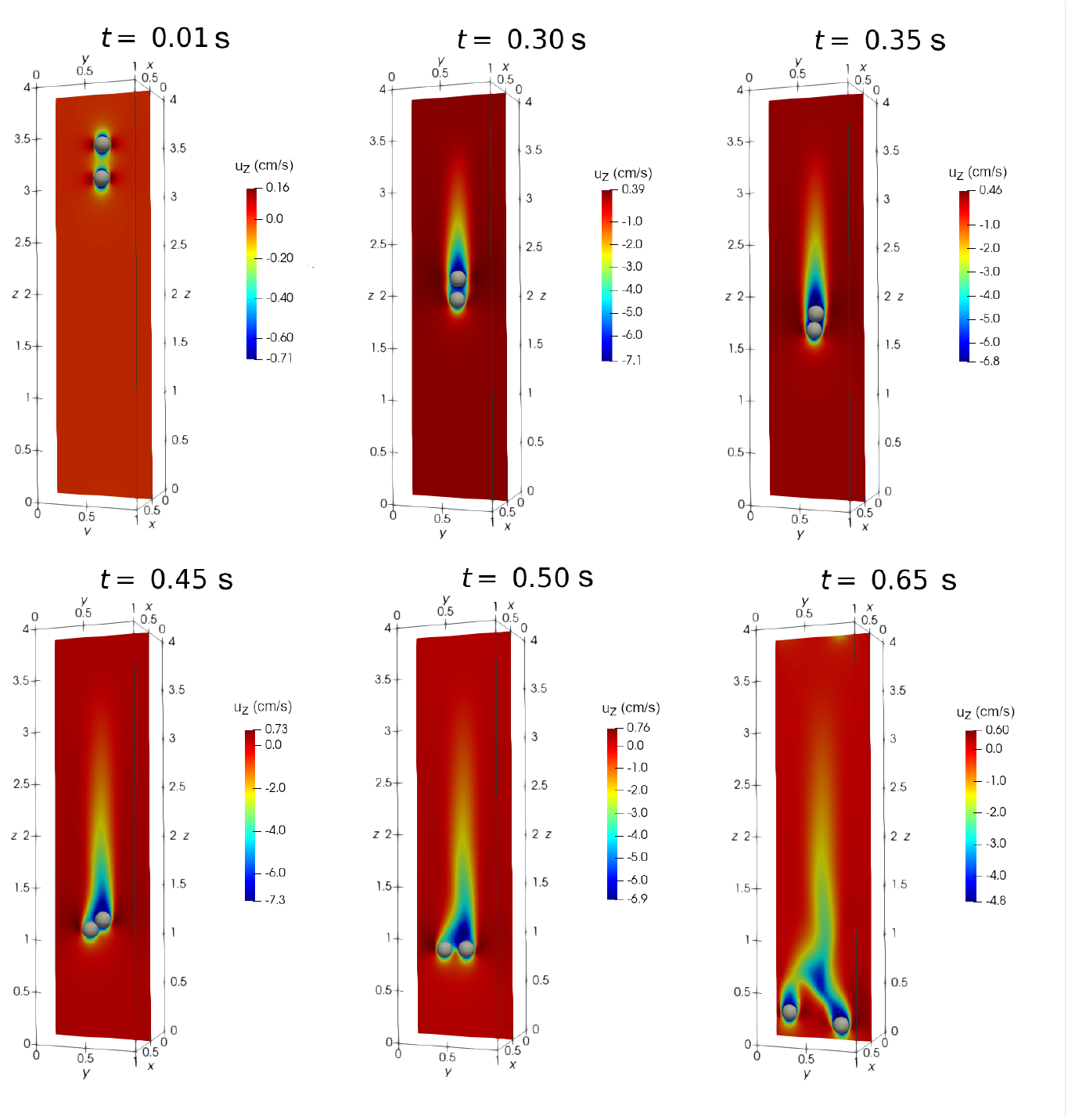}%
\caption[]
{The drafting-kissing-tumbling (DKT) benchmark case study simulated using Algorithm~\ref{alg1} with no magnetic field. The positions of spheres at $t=0.01,~0.30,~0.35,~0.45,~0.50$ and $0.65$ s and the contour of the longitudinal ($z-$component) fluid velocity, $u_z~$(cm/s), at the midplane $x=0.5$~cm are shown.} 
\label{fig:DKTTime}
\end{figure}

When comparing our results with the results computed by \citet{Glowinsky2000} with $h_{\Omega}=1/60$ and $h_{\Omega}=1/80$, it can be seen that they both predicted similar physical behaviors but with small discrepancy on timing. It must be noted that the kissing, drafting, and tumbling (DKT) benchmark case study is a non-smooth case involving several symmetry breaking. The exact agreement between different numerical algorithms after the kissing phenomenon is difficult to achieve, in part due to the lack of achieving mesh-independent results, or the use of different inter-particle contact models that influences the particles' position drastically. To show the completeness of our solution, Fig.~\ref{fig:DKTTime} presents particles location and the contour distribution for the longitudinal fluid velocity, $u_z$~(cm/s), obtained at the middle plane $x=0.5$~cm for times $t=0.01,~0.30,~0.35,~0.45,~0.50$, and $0.65$~s obtained with M2. One can distinctly observe that the drafting ($t=0.3$~s), kissing ($t=0.35$~s), and tumbling ($t=0.45$~s) phenomena are indeed taking place. Next test-cases explore how the DKT benchmark case study changes when particles are magnetized under a constant magnetic field.

\subsection{DKT phenomenon under magnetic field}
\label{sec:CS2}

\begin{figure}[H]
\captionsetup[subfigure]{justification=justified,singlelinecheck=false}
    \centering
    {\renewcommand{\arraystretch}{0}
    \begin{tabular}{c@{}c}
    \begin{subfigure}[b]{.5\columnwidth}
        \centering
        \caption{{}}
        \includegraphics[width=0.6\columnwidth]{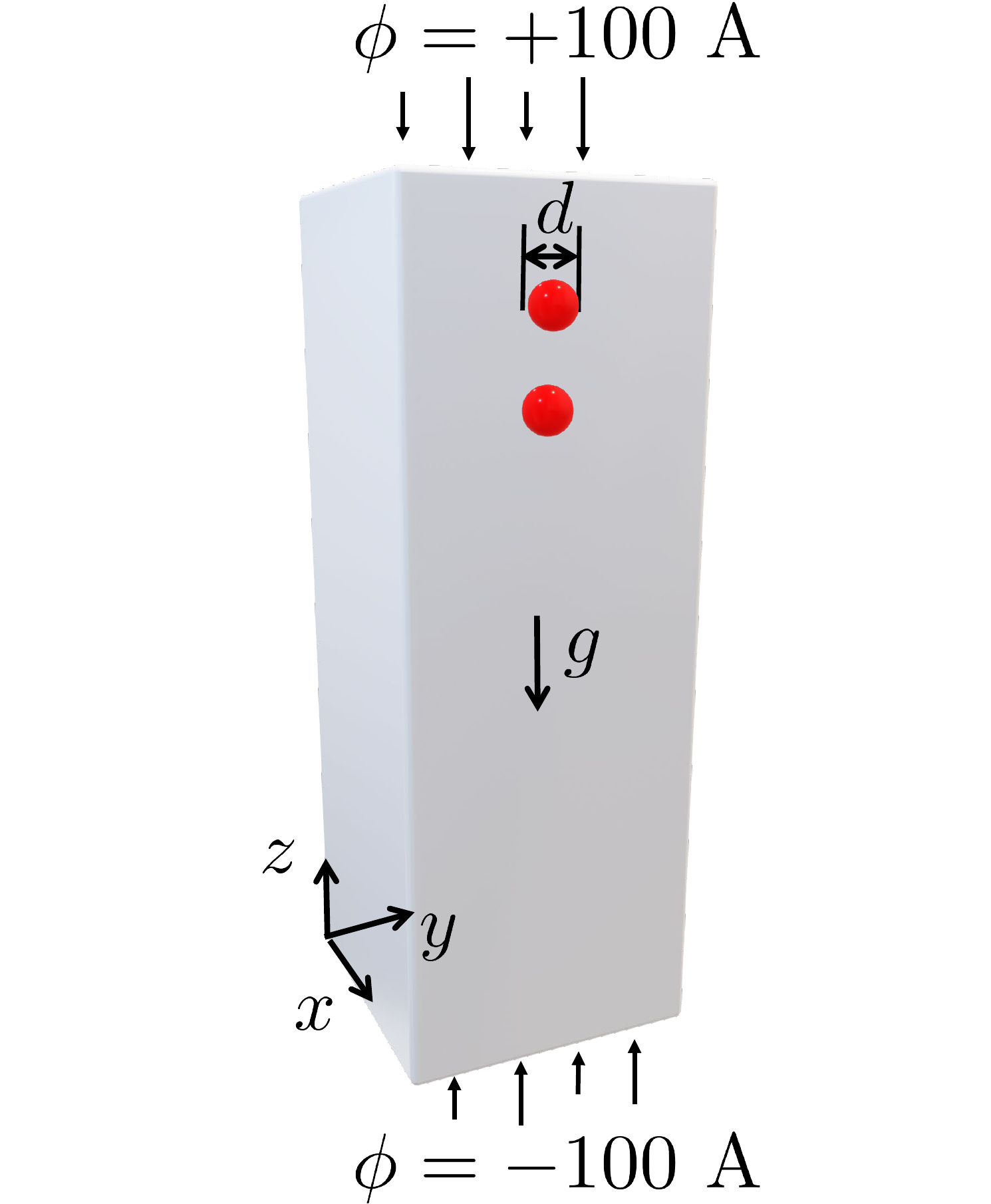}%
        \label{}
    \end{subfigure}\\
        \begin{subfigure}[b]{.5\columnwidth}
        \centering
        \caption{{}}
        \includegraphics[width=0.6\columnwidth]{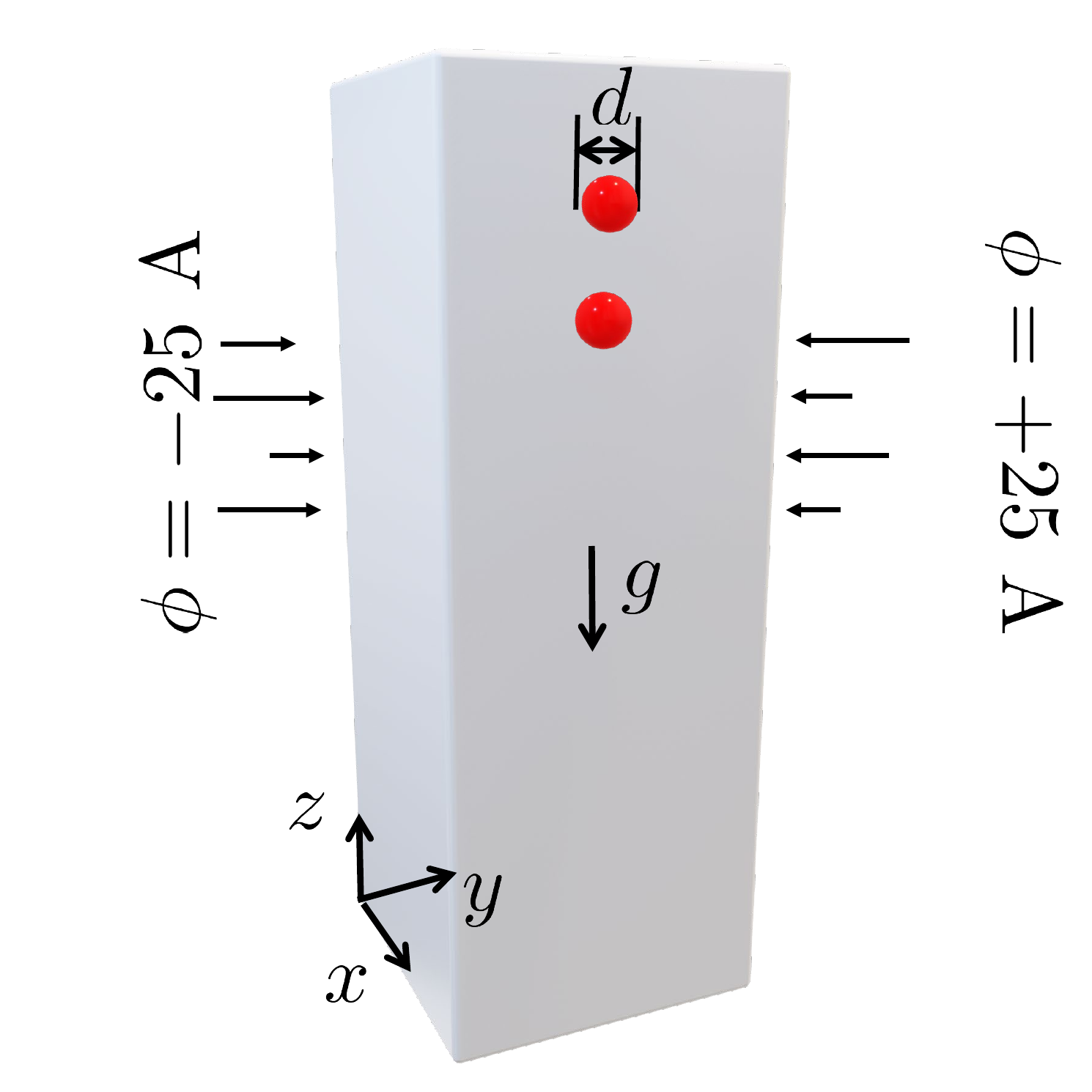}%
        \label{}
    \end{subfigure}\\
    \end{tabular}}
    \caption[]
    {Configuration of the drafting-kissing-tumbling (DKT) benchmark case study with application of an external magnetic field potential in the (a) vertical  ($z$-axis) and (b) horizontal ($y$-axis) directions. The schematic diagram illustrates the computational domain including the coordinate system, the gravitational acceleration $g$, the potential magnetic field, and the initial positions of the spheres  located on $(0.5, 0.5, 3.5)$ and $(0.5, 0.49, 3.16)$.} 
    \label{fig:DKTMagneticVH}
\end{figure}

This computational experiments examines the effect of the application of an external magnetic field on the DKT benchmark case study described in Section~\ref{sec:CS1}. We apply the external magnetic field both vertically (see Fig.~\ref{fig:DKTMagneticVH}(a)) and horizontally  (see Fig.~\ref{fig:DKTMagneticVH}(b)), and explore how the magnetic field affects the sedimentation of the two magnetic spheres, i.e., the DKT phenomenon  \cite{Ke2017}. In both cases, the applied magnetic potential gradient field, $\nabla \phi$, is set to 50 A/m. In addition, the fixed dipole model (see Eq.~(\ref{eq:fixedDipoleM})) is employed to magnetize the particles with a relative susceptibility of $\chi = 2000$ \cite{Ly1999}.

\begin{figure}[H]
\captionsetup[subfigure]{justification=justified,singlelinecheck=false}
    \centering
    {\renewcommand{\arraystretch}{0}
    \begin{tabular}{c@{}c}
    \begin{subfigure}[b]{.5\columnwidth}
        \centering
        \caption{{}}
        \includegraphics[width=1.0\columnwidth]{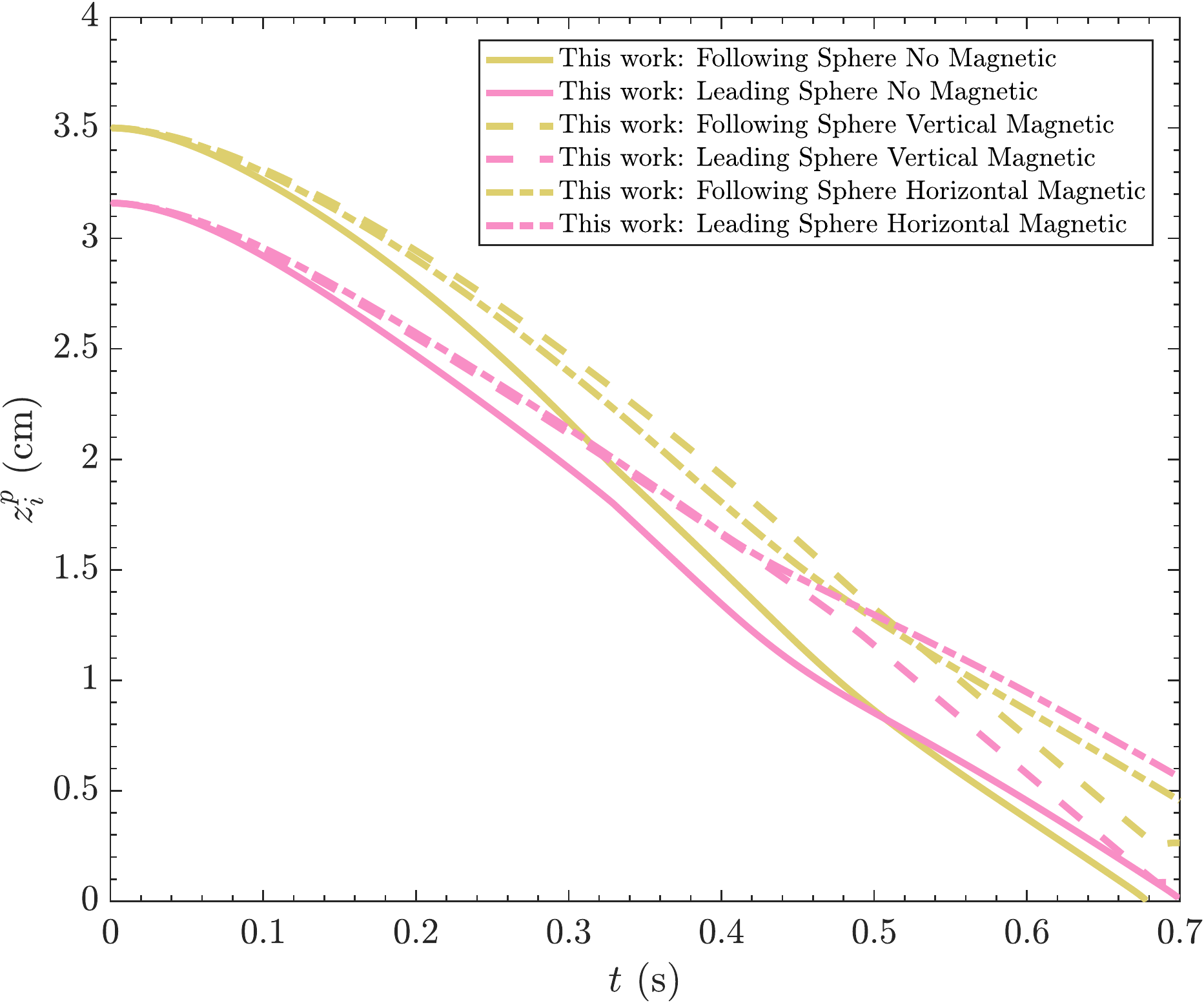}%
        \label{}
    \end{subfigure}\\
        \begin{subfigure}[b]{.5\columnwidth}
        \centering
        \caption{{}}
        \includegraphics[width=1.0\columnwidth]{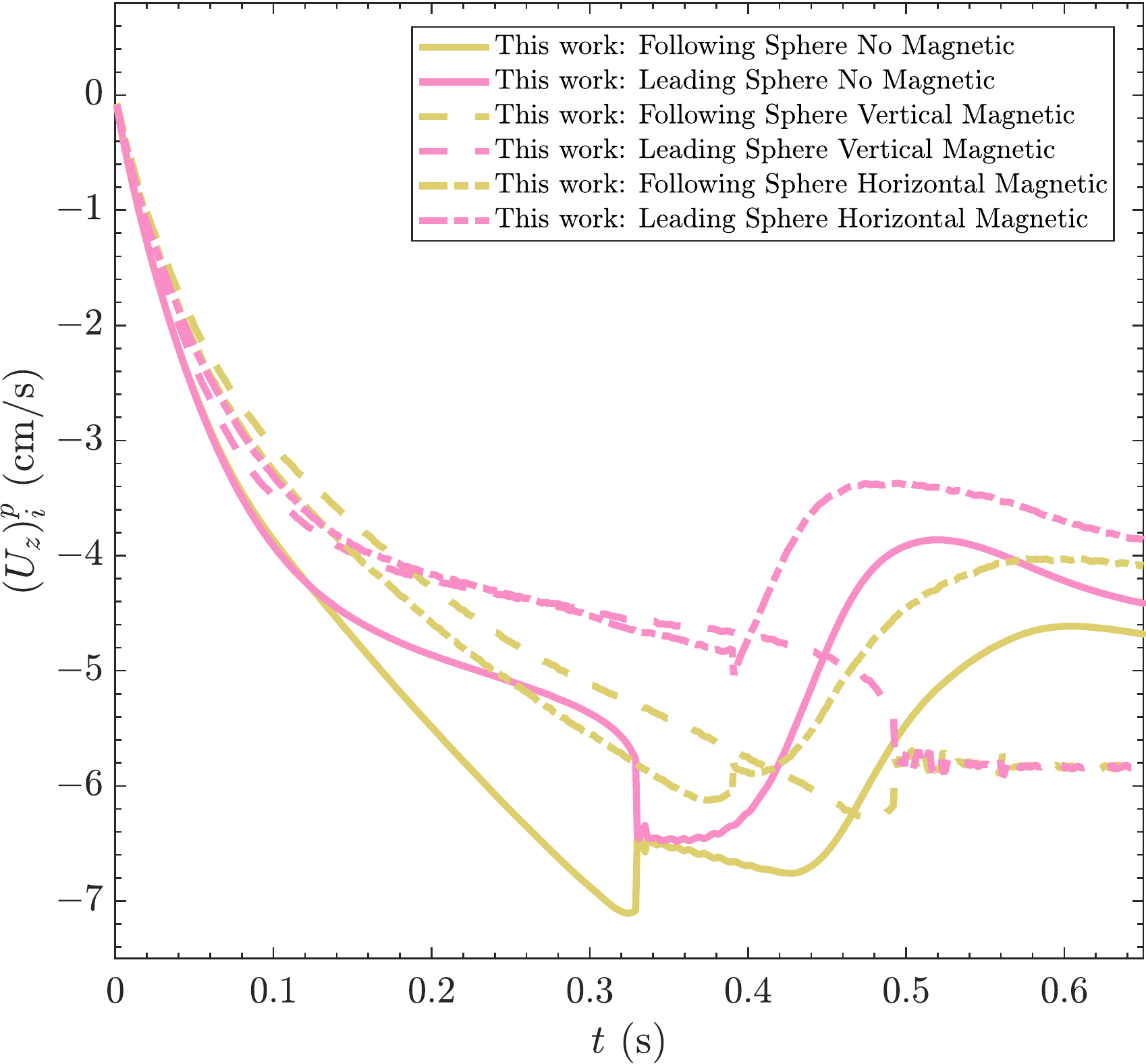}%
        \label{}
    \end{subfigure}\\
    \end{tabular}}
    \caption[]
    {A comparison of the $z$-component of the spheres' (a) center location, and (b) translation velocity as function of time  obtained using Algorithm \ref{alg1} to simulate the drafting-kissing-tumbling (DKT) benchmark case study under no magnetic field, vertical magnetic field, and horizontal magnetic field.} 
    \label{fig:DKTOpenFOAMMagn}
\end{figure}

\begin{figure}[H]
\centering
\includegraphics[width=1.0\columnwidth]{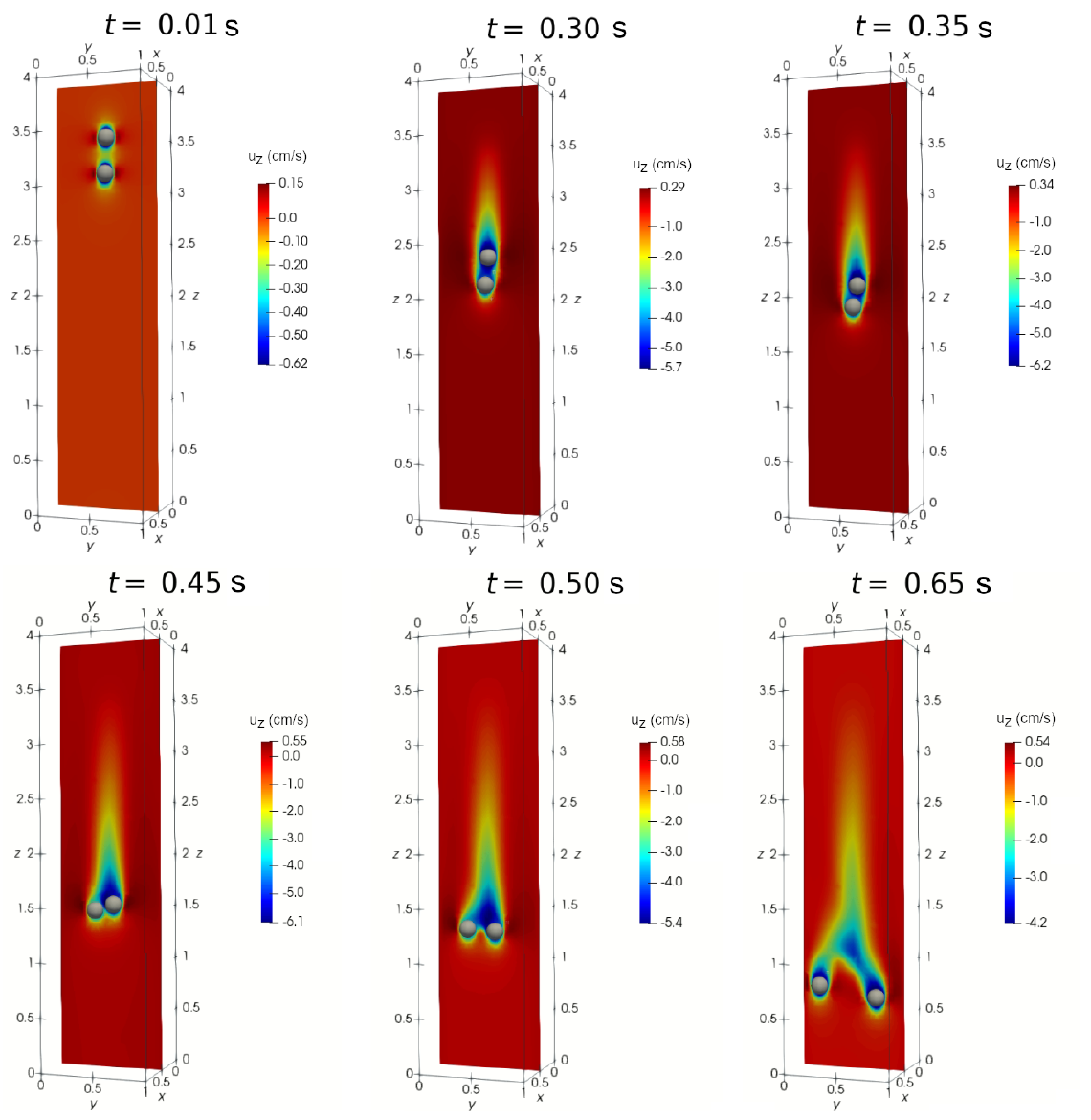}%
\caption[]
{The change in drafting-kissing-tumbling (DKT) benchmark case study under horizontal magnetic field simulated using Algorithm~\ref{alg1}. The positions of spheres at $t=0.01,~0.30,~0.35,~0.45,~0.50$ and $0.65$~s, and the contour of the longitudinal ($z-$component) fluid velocity, $u_z$~(cm/s), at the midplane $x=0.5$~cm are shown.} 
\label{fig:DKTHorizontalTime}
\end{figure}

Figure \ref{fig:DKTOpenFOAMMagn} shows the $z$-component of the spheres centers, $z_i^p$, and of the $z$-component of the spheres translation velocities, $(U_z)_i^p$, as function of time for the calculations using mesh M2. In addition, the results of the DKT benchmark case study under no magnetic field  (obtained in Section~\ref{sec:CS1}) are also shown for comparison purposes. When the external magnetic field is applied in the vertical direction, the two magnetic particles are attracted together forming a string ($t\approx 0.5~$s). The string last until they contact the bottom wall of the domain. On the other turn, when the external magnetic field is induced in the horizontal direction, the spherical particles do not approach each other, but instead they tumble side-by-side. During the rest of the sedimentation process, the wake generated by the leading particle leads to a faster settling of the following particle ($t > 0.5~$s), also see Fig.~\ref{fig:DKTHorizontalTime} for an illustrative representation of this phenomenon.

\begin{figure}[H]
\centering
\includegraphics[width=1.0\columnwidth]{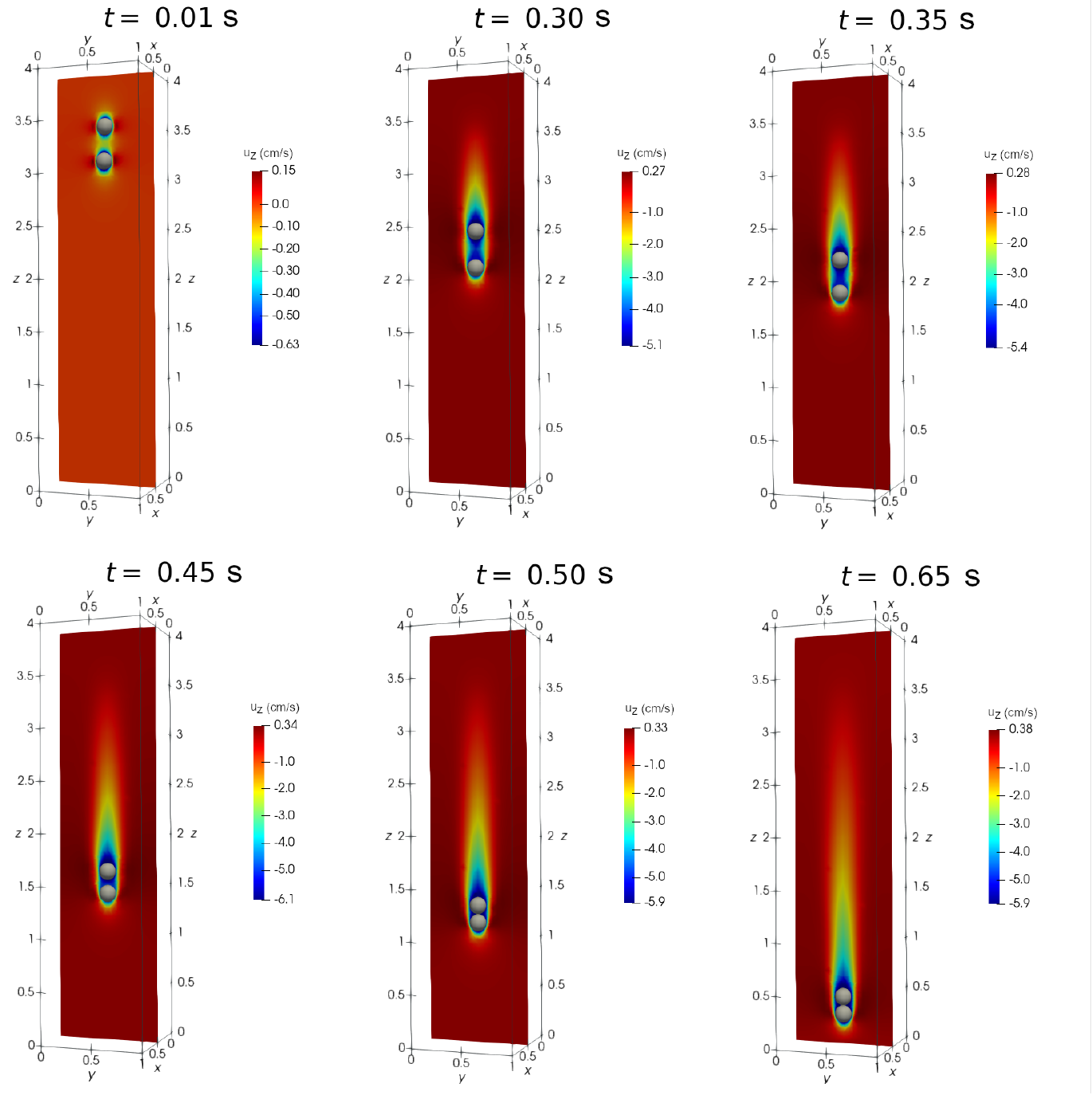}%
\caption[]
{The change in drafting-kissing-tumbling (DKT) benchmark case study under vertical magnetic field simulated using Algorithm~\ref{alg1}. The positions of spheres at $t=0.01,~0.30,~0.35,~0.45,~0.50$ and $0.65$ s, and the contour of the longitudinal ($z-$component) fluid velocity, $u_z~$(cm/s), at the midplane $x=0.5$~cm are shown.} 
\label{fig:DKTVerticalTime}
\end{figure}

Figure \ref{fig:DKTVerticalTime} presents particles settling under vertical magnetic field. The particle location and the contour distribution for the longitudinal fluid velocity, $u_z~$(cm/s),  obtained at the midplane $x=0.5$~cm for times $t=0.01,~0.30,~0.35,~0.45,~0.50$, and $0.65$ s are shown. As can be seen, the particles experience longer drafting period, and form a tight string that does not get separated in the rest of the sedimentation process. 

Figure \ref{fig:DKTHorizontalTime} shows the settling of the spherical particles under an horizontal magnetic field. The particle location and the contour distribution for the longitudinal fluid velocity, $u_z$~(cm/s) obtained at the midplane $x=0.5$~cm for times $t=0.01,~0.30,~0.35,~0.45,~0.50$, and $0.65$~s are shown in Fig.~\ref{fig:DKTHorizontalTime}. In this case, the direction of the particles sedimentation is transverse to the magnetic field direction, and hence, particles form a repulsive magnetic force \cite{Ke2017}. For that reason, the particles, instead of approaching and contacting each other, just tumble as a non-kissing pair ($t\approx 0.50$~s). Shortly after the tumble, the particles approach the vertical walls, where the external magnetic field is applied ($t\approx 0.65$~s). 

\subsection{Multi-particle chaining under magnetic field: 2D}
\label{sec:CS3}

\begin{figure}[H]
\captionsetup[subfigure]{justification=justified,singlelinecheck=false}
    \centering
    {\renewcommand{\arraystretch}{0}
    \begin{tabular}{c@{}c}
    \begin{subfigure}[b]{0.7\columnwidth}
        \centering
        \caption{{}}
        \includegraphics[width=0.6\columnwidth]{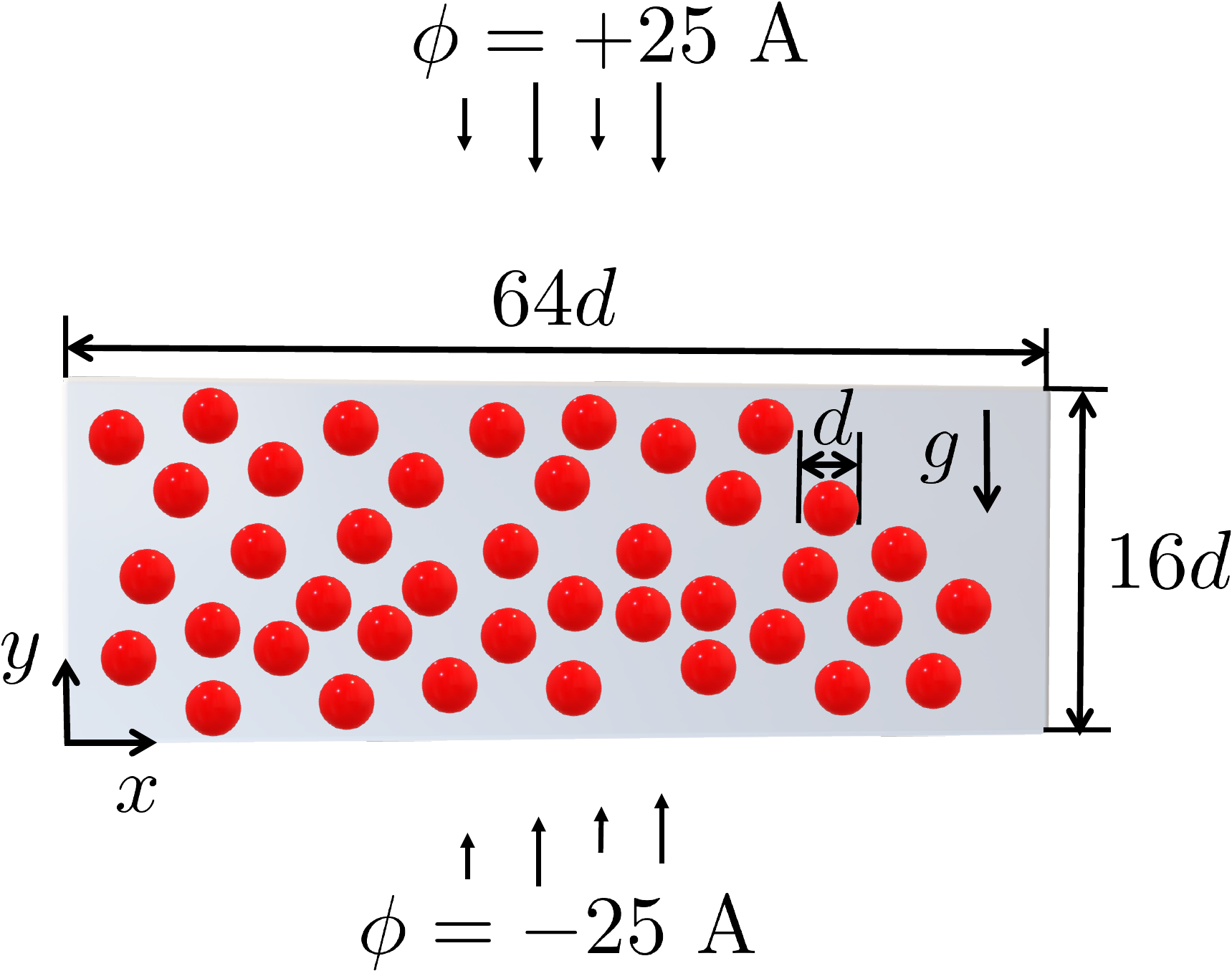}%
        \label{}
    \end{subfigure}\\
        \begin{subfigure}[b]{0.7\columnwidth}
        \centering
        \caption{{}}
        \includegraphics[width=0.6\columnwidth]{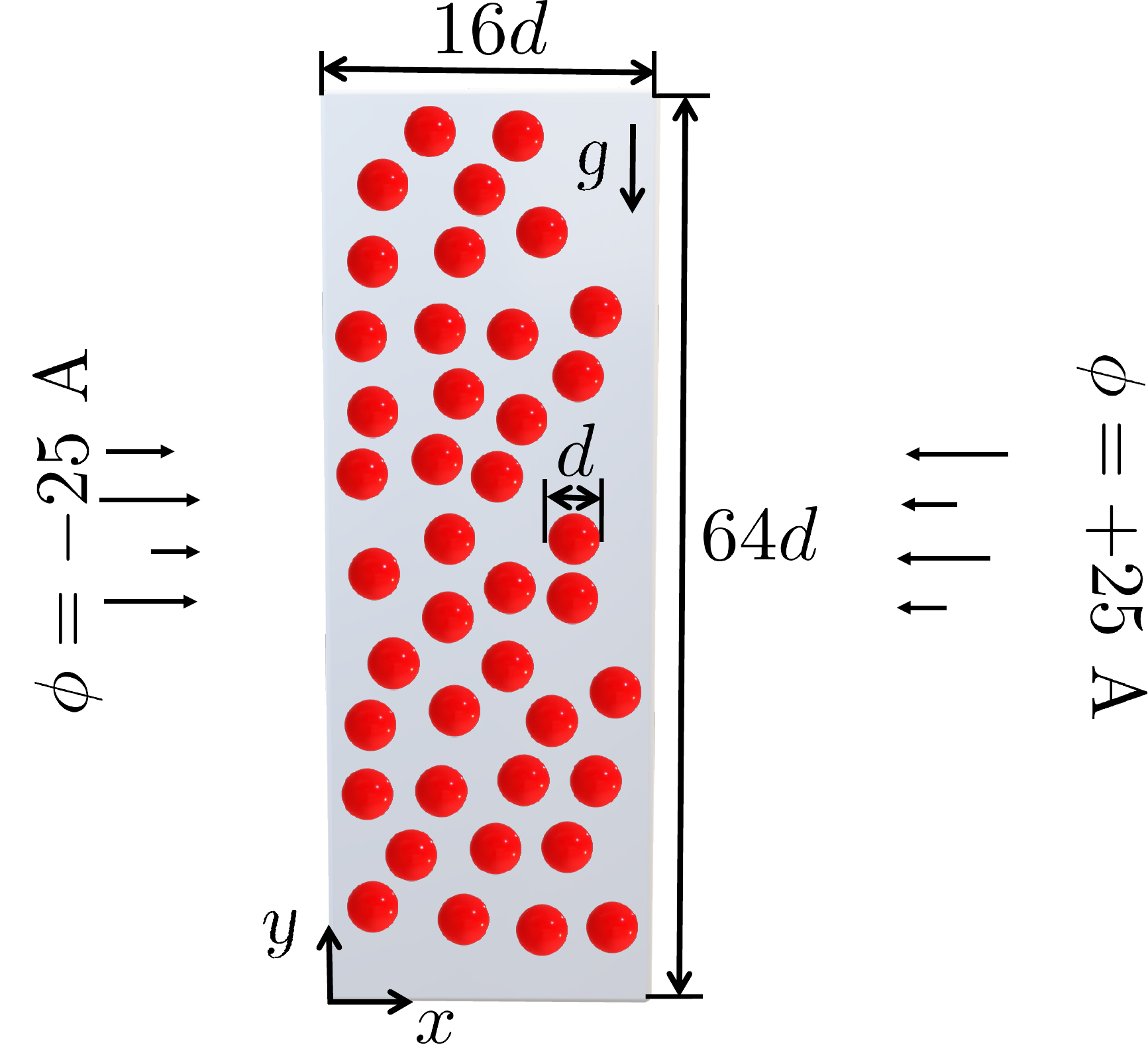}%
        \label{}
    \end{subfigure}\\
    \end{tabular}}
    \caption[]
    {Configuration of the multi-particle problem where spheres centers are located on a 2D plane and with application of an external magnetic potential field in the (a) vertical ($y$-axis) and (b) horizontal ($x$-axis) directions. The schematic diagram illustrates the computational domain including the coordinate system, the potential magnetic field, the gravitational acceleration $g$, and random position of particles.} 
    \label{fig:MultiPart2DMagneticVH}
\end{figure}

In this test case, we analyze the motion of a random array of magnetic particles whose centers are located at the midplane of a rectangular box filled with a Newtonian fluid under the influence of external magnetic fields \cite{Han2010, Ke2017, Ly1999}. Two computational domains are employed as $\Omega_h = \left[0,4\right]\times\left[0,1\right]\times \left[0,1\right]$~cm$^3$ and $\Omega_v = \left[0,1\right]\times\left[0,4\right]\times \left[0,1\right]$~cm$^3$  (see Fig.~\ref{fig:MultiPart2DMagneticVH}). The initial positions of the spheres centers, with diameter $d=1/16$~cm and density of $\rho_s = 1.01$~g/cm$^3$, are randomly generated and constrained such that the minimum distance between particles and between the particles and walls is equal to $1.5d$. The spheres move under the action of gravity, hydrodynamic forces, mutual dipole-dipole forces, and the applied external magnetic force \cite{Ke2017}. 

Two area fractions of spheres were tested, $20\%$ and $30\%$, corresponding to 260 and 390 spheres under the effect of both a vertical and horizontal magnetic fields with magnetic potential gradient of $\nabla\phi=50$~A/m. The fluid and the spheres are initially at rest. On the channel walls, the no-slip boundary condition is applied for the fluid velocity. A cyclic boundary condition is applied on the other boundaries. In addition, the fluid density and kinematic viscosity are set to $\rho_f = 1$~g/cm$^3$ and $\nu = 0.01$~cm$^2$/s, respectively. The dipole-dipole magnetic forces and torques are calculated using the mutual dipole model, see Eq.~(\ref{eq:mutualDipole}), with a relative susceptibility of $\chi = 2000$ \cite{Ly1999}. For the inter-particle contacts and particle-wall contacts, the coefficient of normal restitution, coefficient of friction, Poisson's ratio and  Young's modulus are considered to be $0.90$, $0.33$, $0.33$, and  $7\times 10^8$ Pa, respectively \cite{Ke2017}. 

The calculations were performed in an hexahedral mesh with initial configuration $128\times 32\times 32$ grid cells for the horizontal domain (M$_h$) and $32\times 128\times 32$ grid cells for the vertical domain (M$_v$). Again, the dynamic mesh refinement was employed in the calculations with two levels of refinement. The time-step used in the simulations is $\Delta t = 10^{-4}$~s corresponding to a maximum Courant number of $0.1$. The total computational elapsed time for the simulations was approximately 2h05m and 2h35m for the $20\%$ and $30\%$ particle's area fractions executed on a 3.00-GHz 48 cores Intel Xeon Gold 6248R CPU processor with 128 GB of RAM.

\begin{figure}[H]
\centering
\includegraphics[width=1.0\columnwidth]{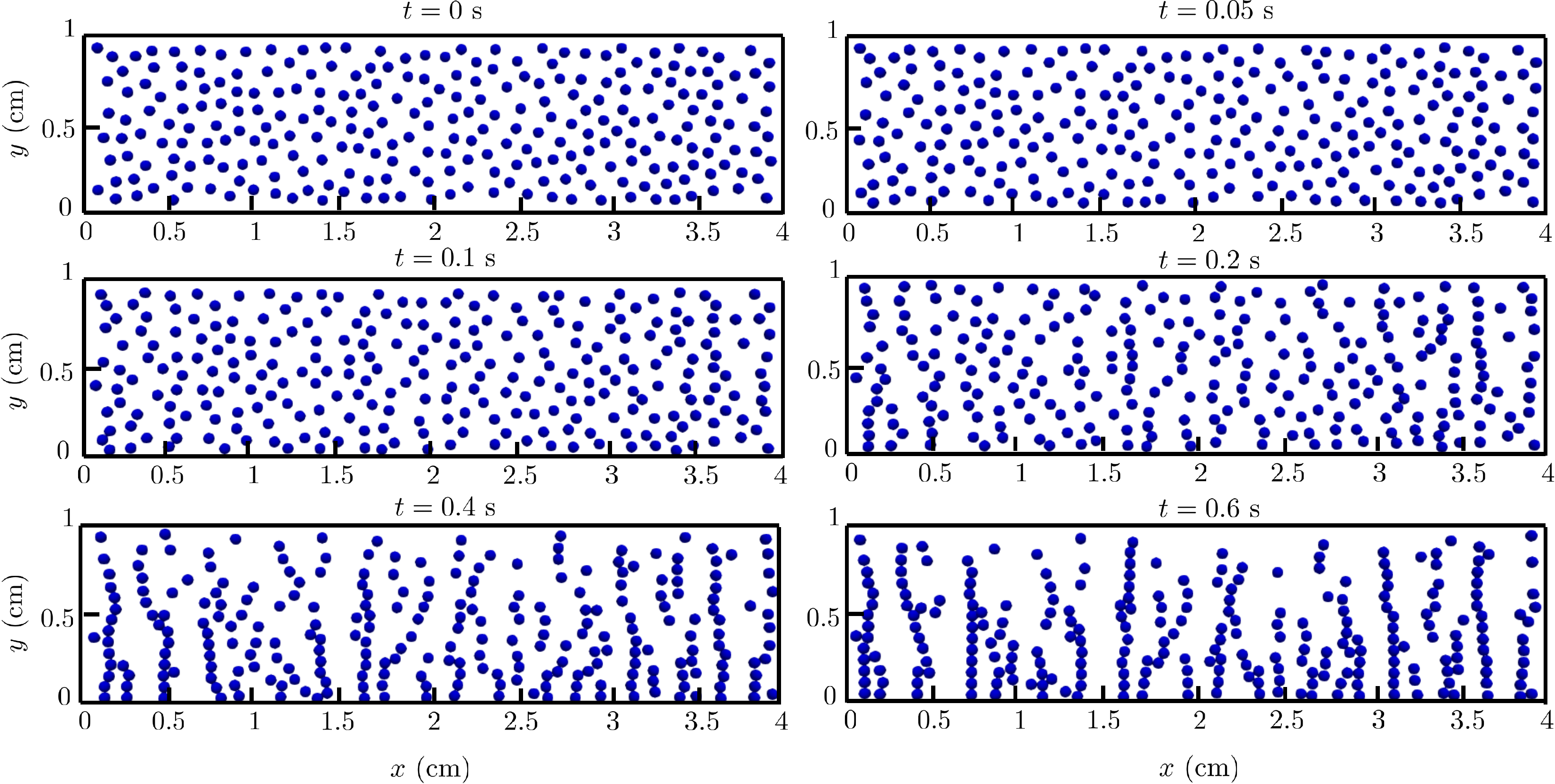}%
\caption[]
{Behavior of a random array of magnetic spheres on a 2D domain with $20\%$ particle area fraction at $t=0,~0.05,~0.1,~0.2,~0.4$ and $0.6$ s under the action of gravity and an external magnetic field applied in the vertical direction.}
\label{fig:RandomArraysHorizontal20}
\end{figure}

\begin{figure}[H]
\centering
\includegraphics[width=1.0\columnwidth]{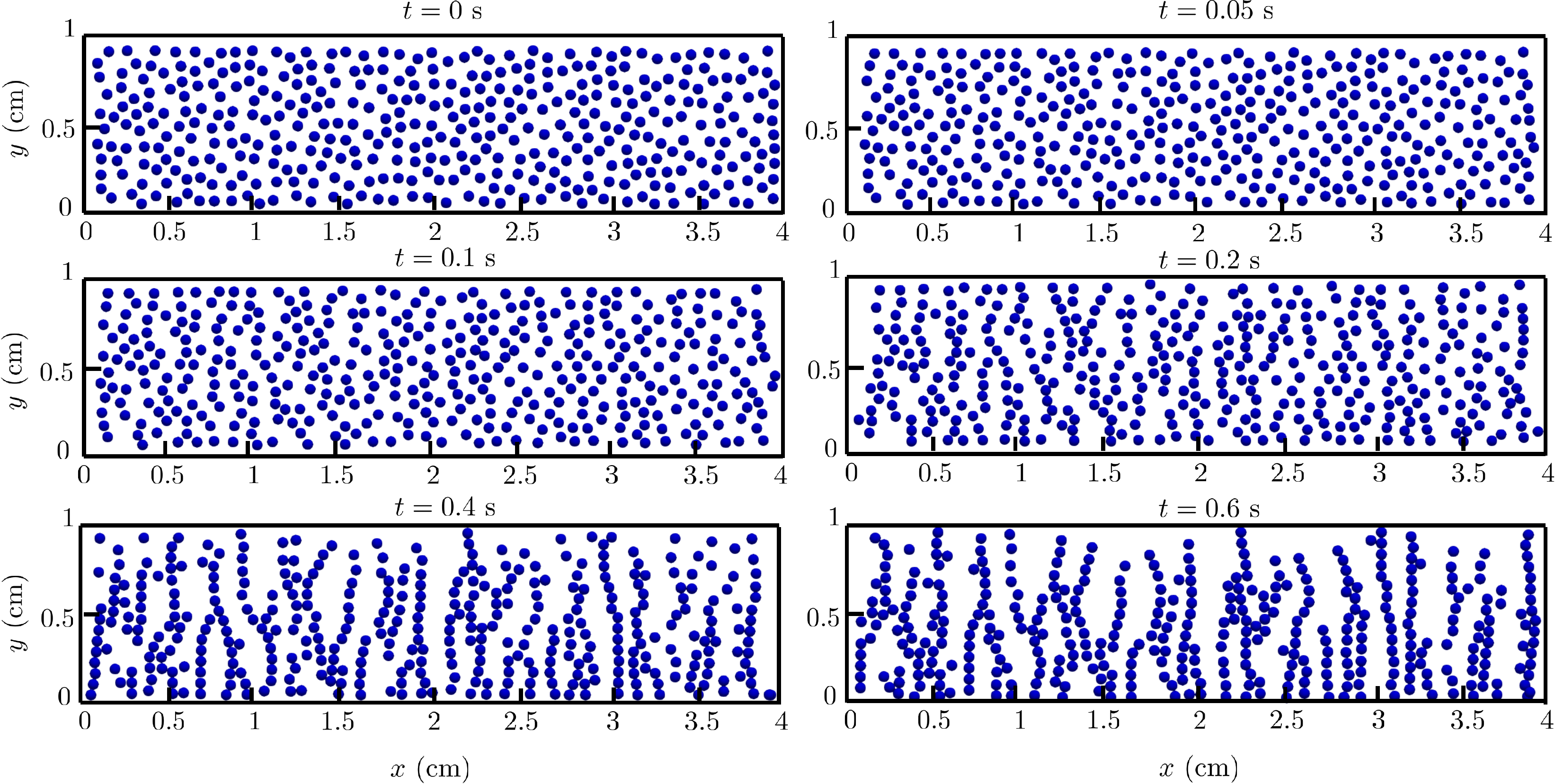}%
\caption[]
{Behavior of a random array of magnetic spheres on a 2D domain with $30\%$ particle area fraction at $t=0,~0.05,~0.1,~0.2,~0.4$ and $0.6$ s under the action of gravity and an external magnetic field applied in the vertical direction.}
\label{fig:RandomArraysHorizontal30}
\end{figure}

Figures~\ref{fig:RandomArraysHorizontal20} and \ref{fig:RandomArraysHorizontal30} show the snapshots of 260 and 390 particles moving in the rectangular channel under the effect of gravity and an external magnetic field applied in the vertical direction. At the initial instants of the simulations ($t\leq 0.2$~s), short fragmented chains or clusters of particles are formed in the $y-$direction (the same as the applied external magnetic field direction). Subsequently, at later instants ($0.4 \leq t\leq 0.6$~s), the short chains start to merge together and form long chains, i.e., they form mesoscale structures made of magnetic particles with shapes and orientations comparable to the results presented by \citet{Han2010}, \citet{Ke2017} and \citet{Ly1999}.


\begin{figure}[H]
\centering
\includegraphics[width=1.0\columnwidth]{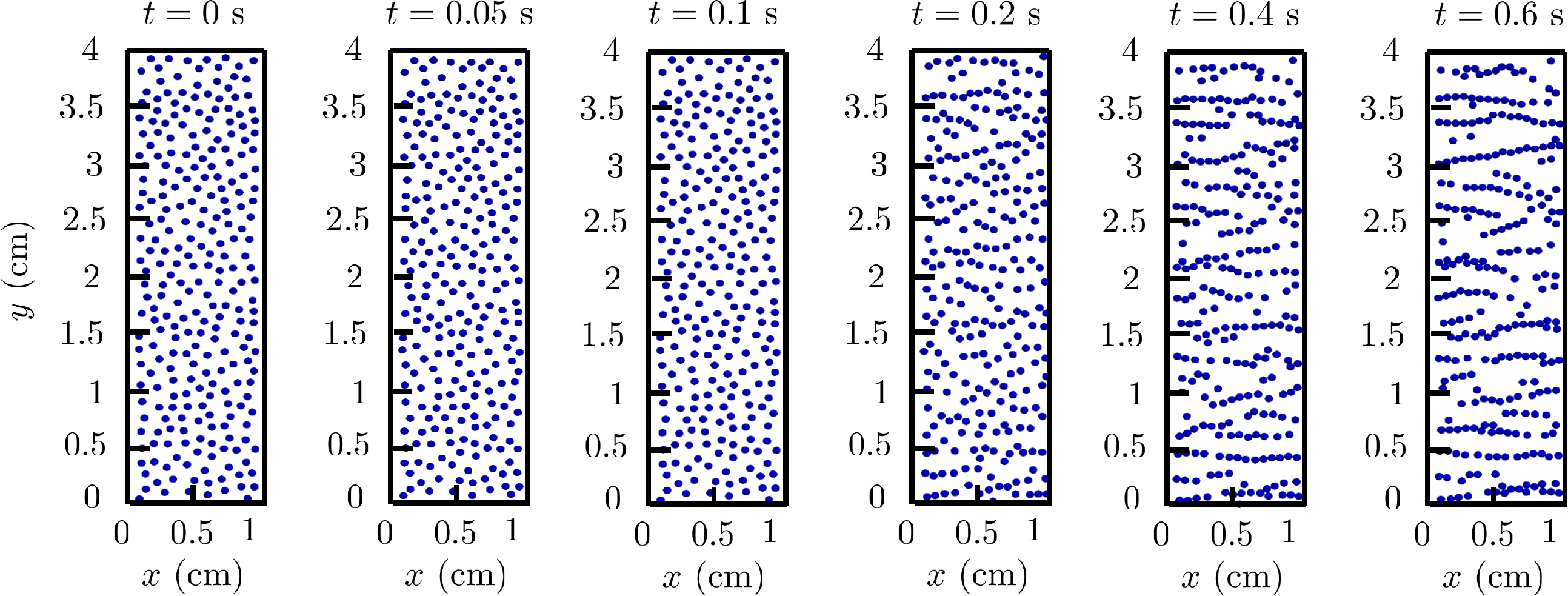}%
\caption[]
{Behavior of a random array of magnetic spheres on a 2D domain with $20\%$ particle area fraction at $t=0,~0.05,~0.1,~0.2,~0.4$ and $0.6$ s under the action of gravity and an external magnetic field applied in the horizontal direction.}
\label{fig:RandomArraysVertical20}
\end{figure}

\begin{figure}[H]
\centering
\includegraphics[width=1.0\columnwidth]{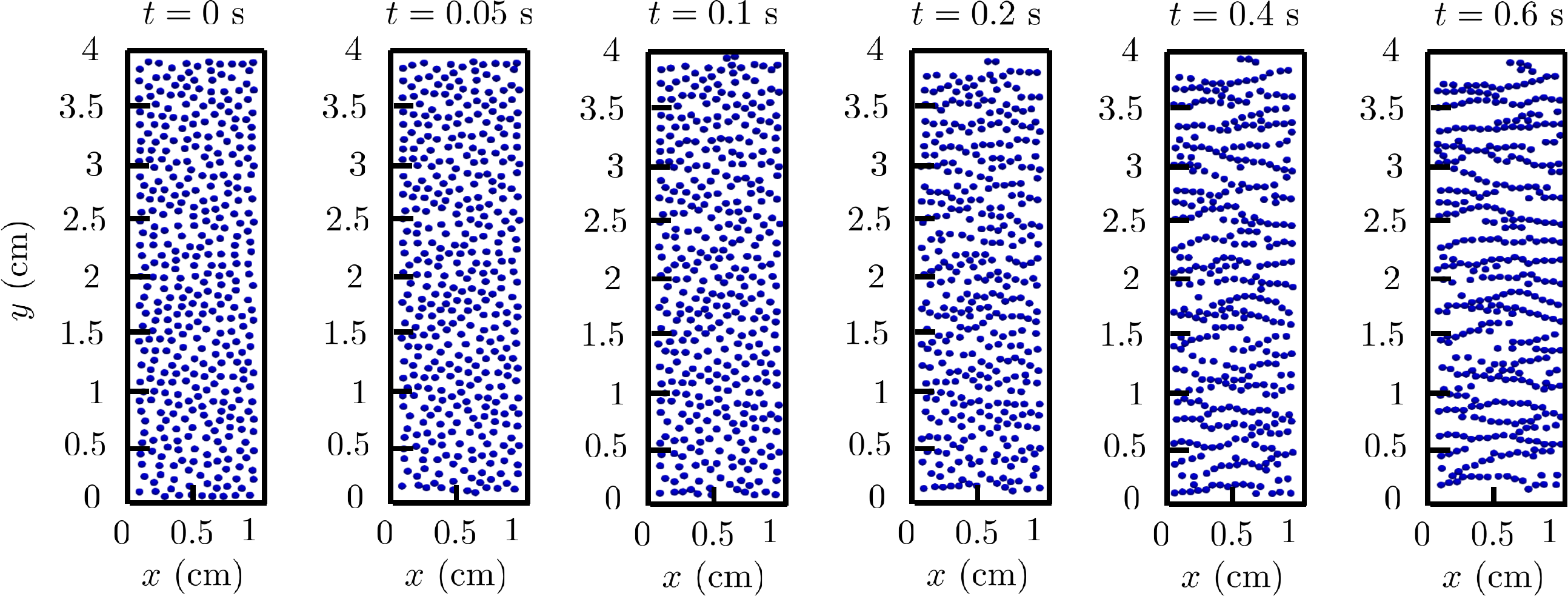}%
\caption[]
{Behavior of a random array of magnetic spheres on a 2D domain with $30\%$ particle area fraction at $t=0,~0.05,~0.1,~0.2,~0.4$ and $0.6$ s under the action of gravity and an external magnetic field applied in the horizontal direction.}
\label{fig:RandomArraysVertical30}
\end{figure}

Figures~\ref{fig:RandomArraysVertical20} and \ref{fig:RandomArraysVertical30} show the snapshots of 260 and 390 particles moving in the rectangular channel under the action of gravity and of an external magnetic field applied in the horizontal direction. Again, at the initial instants of the simulations ($t\leq 0.2$~s), short fragmented chains or clusters of particles are formed in the $x-$direction (the same as the applied external magnetic field direction). Then, at later instants ($0.4 \leq t\leq 0.6$~s), the short chains start to merge together and form long horizontally aligned chains, i.e., they form mesoscale structures made of magnetic particles with distinct shapes and orientations. 

Figures~\ref{fig:RandomArraysHorizontal20} to \ref{fig:RandomArraysVertical30} also show the presence of isolated magnetic particles and a number of chains with shorter lengths. Predominantly, these chains are linear as head-to-tail aggregation of magnetic dipoles, but as claimed by \citet{Ke2017}, \citet{Mohebi1996} and \citet{Fermigier1992}, it is also observed thick particle clusters due to the lateral merging of the linear chains.

\subsection{Multi-particle chaining under magnetic field: 3D}
\label{sec:CS4}

\begin{figure}[H]
\captionsetup[subfigure]{justification=justified,singlelinecheck=false}
    \centering
    {\renewcommand{\arraystretch}{0}
    \begin{tabular}{c@{}c}
    \begin{subfigure}[b]{0.7\columnwidth}
        \centering
        \caption{{}}
        \includegraphics[width=0.6\columnwidth]{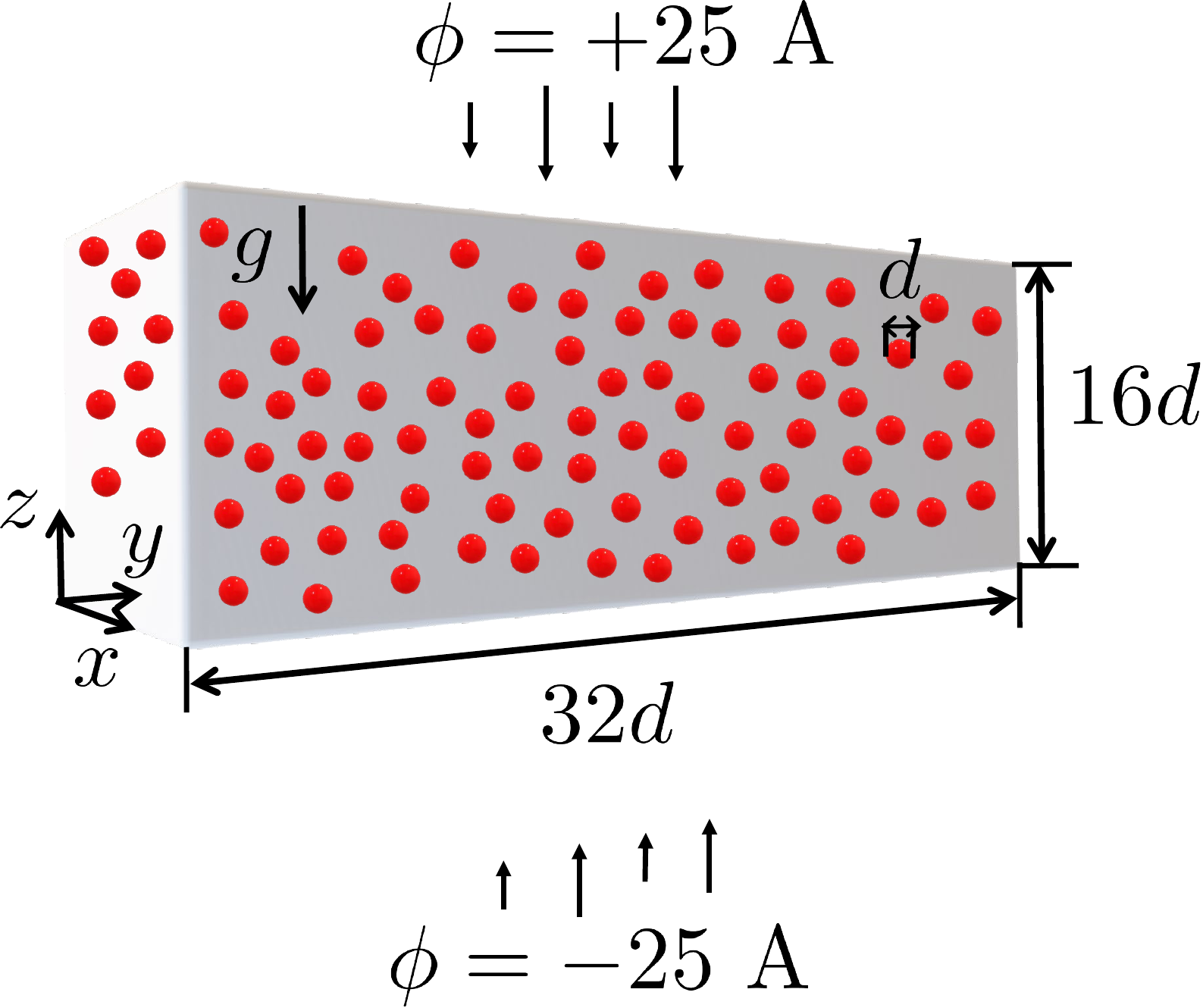}%
        \label{}
    \end{subfigure}\\
        \begin{subfigure}[b]{0.7\columnwidth}
        \centering
        \caption{{}}
        \includegraphics[width=0.7\columnwidth]{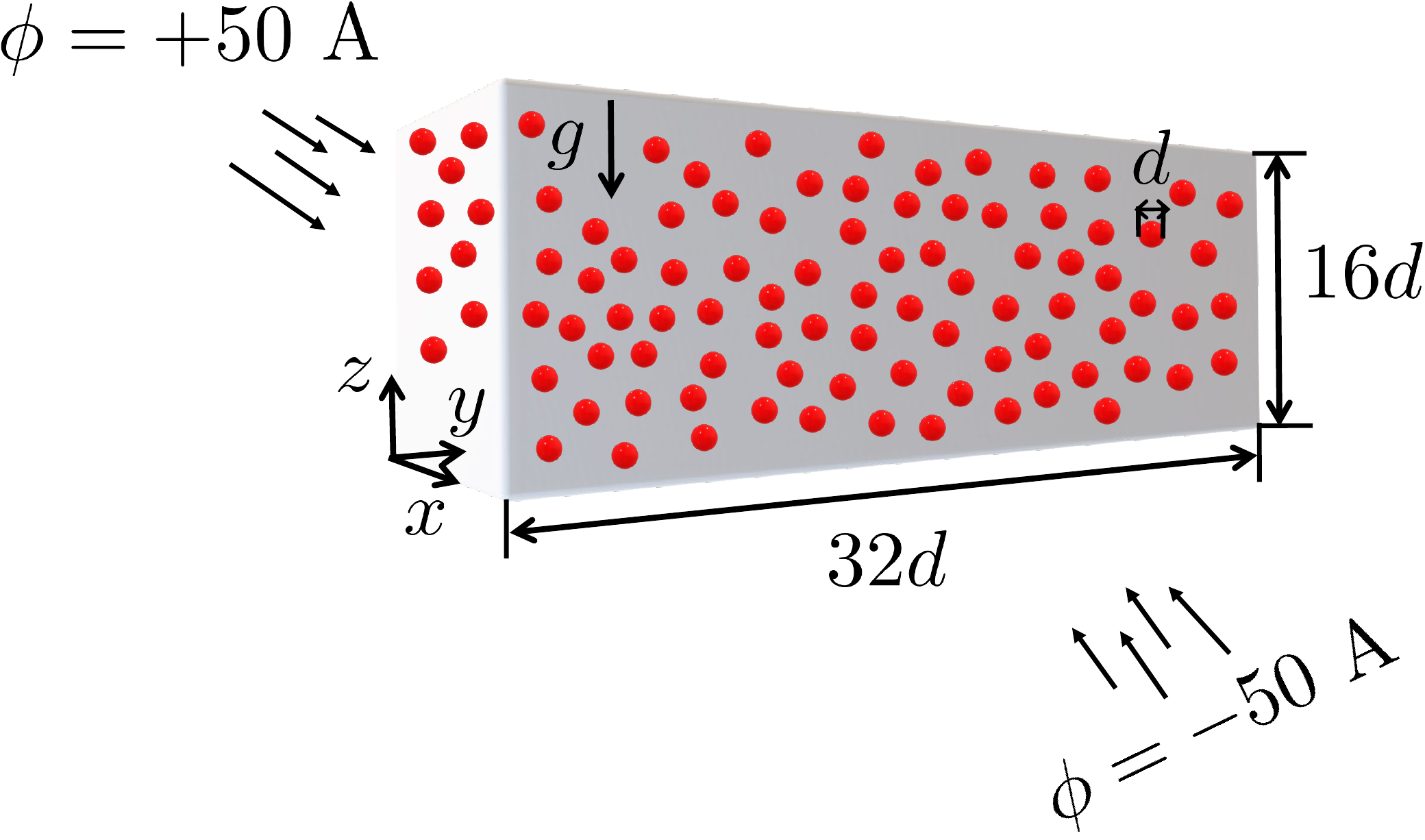}%
        \label{}
    \end{subfigure}\\
    \end{tabular}}
    \caption[]
    {Configuration of the multi-particle problem where spheres centers are located on the 3D spatial domain and with application of an external magnetic potential field in the (a) vertical ($z$-axis) and (b) horizontal ($x$-axis) directions. The schematic diagram illustrates the computational domain including the coordinate system, the potential magnetic field, the gravitational acceleration $g$, and random position of particles.} 
    \label{fig:MultiPart3DMagneticVH}
\end{figure}

In this subsection, we analyze the robustness of the proposed FVM-IBM-DEM-MAG solver by studying the  chain formation in MRFs within a three-dimensional (3D) domain. A random array of magnetic spheres is placed in a rectangular box filled with a Newtonian fluid under the influence of gravity and external magnetic fields applied in different directions \cite{Han2010}. The computational domain employed was $\Omega = \left[0,2\right]\times\left[0,2\right]\times \left[0,1\right]$~cm$^3$ (see Fig.~\ref{fig:MultiPart3DMagneticVH}). The diameter of the spheres is $d=1/16$~cm. The initial positions of the spheres centers are randomly generated with a restriction such that the minimum distance between  particles and between the particles and  walls is higher than $1.5d$. The spheres move under the action of gravity, hydrodynamic forces, mutual dipole-dipole forces, and the applied external magnetic force. The sphere volume fraction was fixed at $1.85\%$, corresponding to 580 spheres. We considered an external magnetic field with magnetic gradient potential $\nabla\phi=50$~A/m applied vertically or horizontally. The fluid and the spheres are initially at rest. On the channel walls, the no-slip boundary condition is imposed for the fluid velocity. A cyclic boundary condition is applied on the other boundaries. The fluid and particle densities are  $\rho_f = 1$~g/cm$^3$ and $\rho_s = 1.01$~g/cm$^3$, respectively. The fluid kinematic viscosity is $\nu = 0.01$~cm$^2$/s. The dipole-dipole magnetic forces and torques are calculated using the mutual dipole model, see Eq.~(\ref{eq:mutualDipole}), with a relative susceptibility of $\chi = 2000$ \cite{Ly1999}. For the inter-particle contacts and particle-wall contacts, the coefficient of normal restitution, coefficient of friction, Poisson's ratio and  Young's modulus are considered to be $0.90$, $0.33$, $0.33$, and  $7\times 10^8$ Pa, respectively \cite{Ke2017}. 

The calculations were performed in an hexahedral mesh with initial configuration of $64\times 64\times 32$ grid cells. Again, the dynamic mesh refinement was employed in the calculations with maxRefinement = 2. The time-step used in the simulations is $\Delta t = 10^{-4}$~s, corresponding to a maximum Courant number of $0.1$. The total computational elapsed time for the simulations was approximately 18h12m executed on a 3.00-GHz 48 cores Intel Xeon Gold 6248R CPU processor with 128 GB of RAM.

\begin{figure}[H]
\centering
\includegraphics[width=1.0\columnwidth]{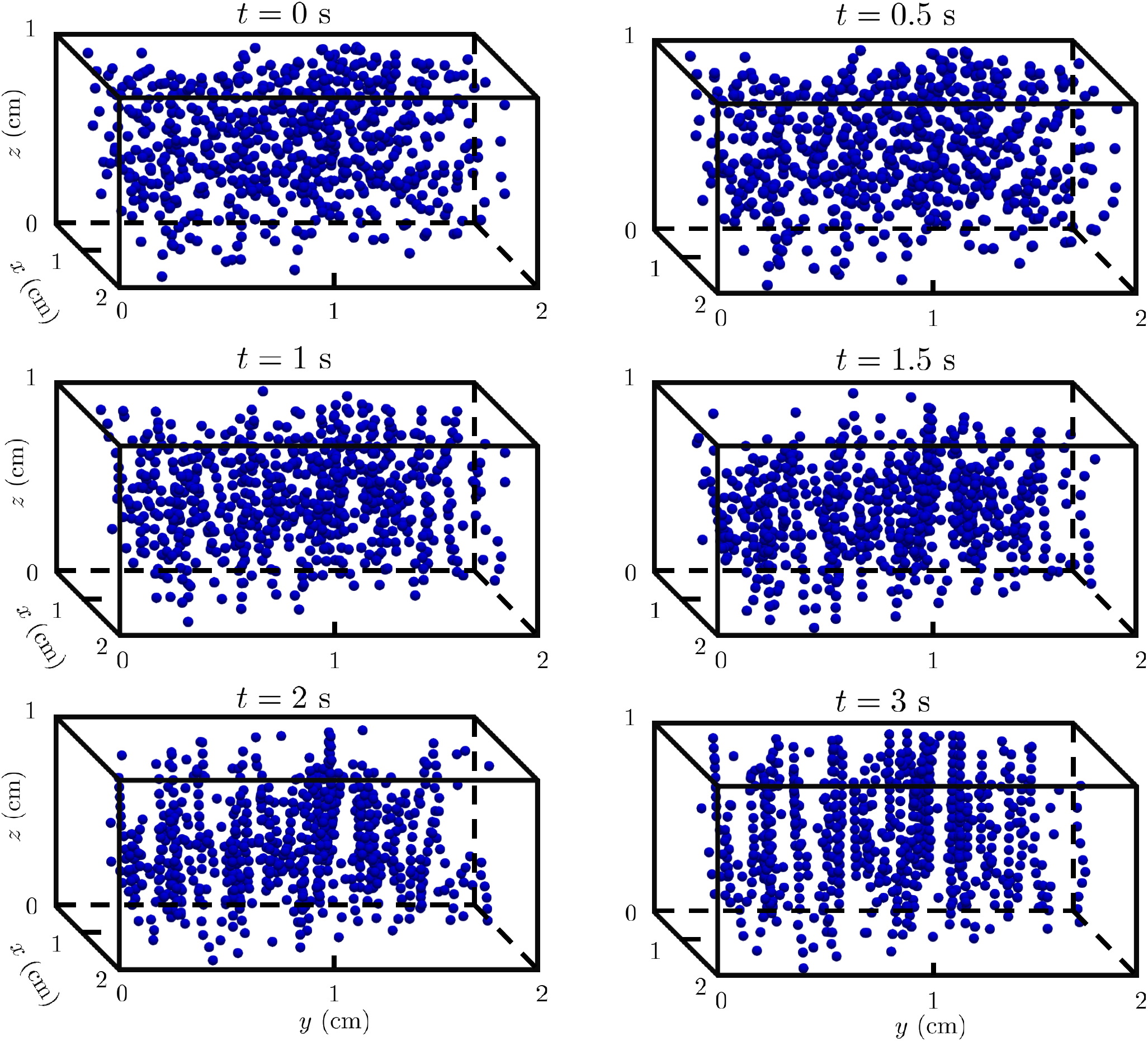}%
\caption[]
{Behavior of a random array of magnetic spheres in a 3D domain with $1.85\%$ particle volume fraction at $t=0,~0.5,~1,~1.5,~2$ and $3$ s under the action of gravity and an external magnetic field applied in the vertical direction.}
\label{fig:RandomArraysVertical}
\end{figure}

\begin{figure}[H]
\centering
\includegraphics[width=1.0\columnwidth]{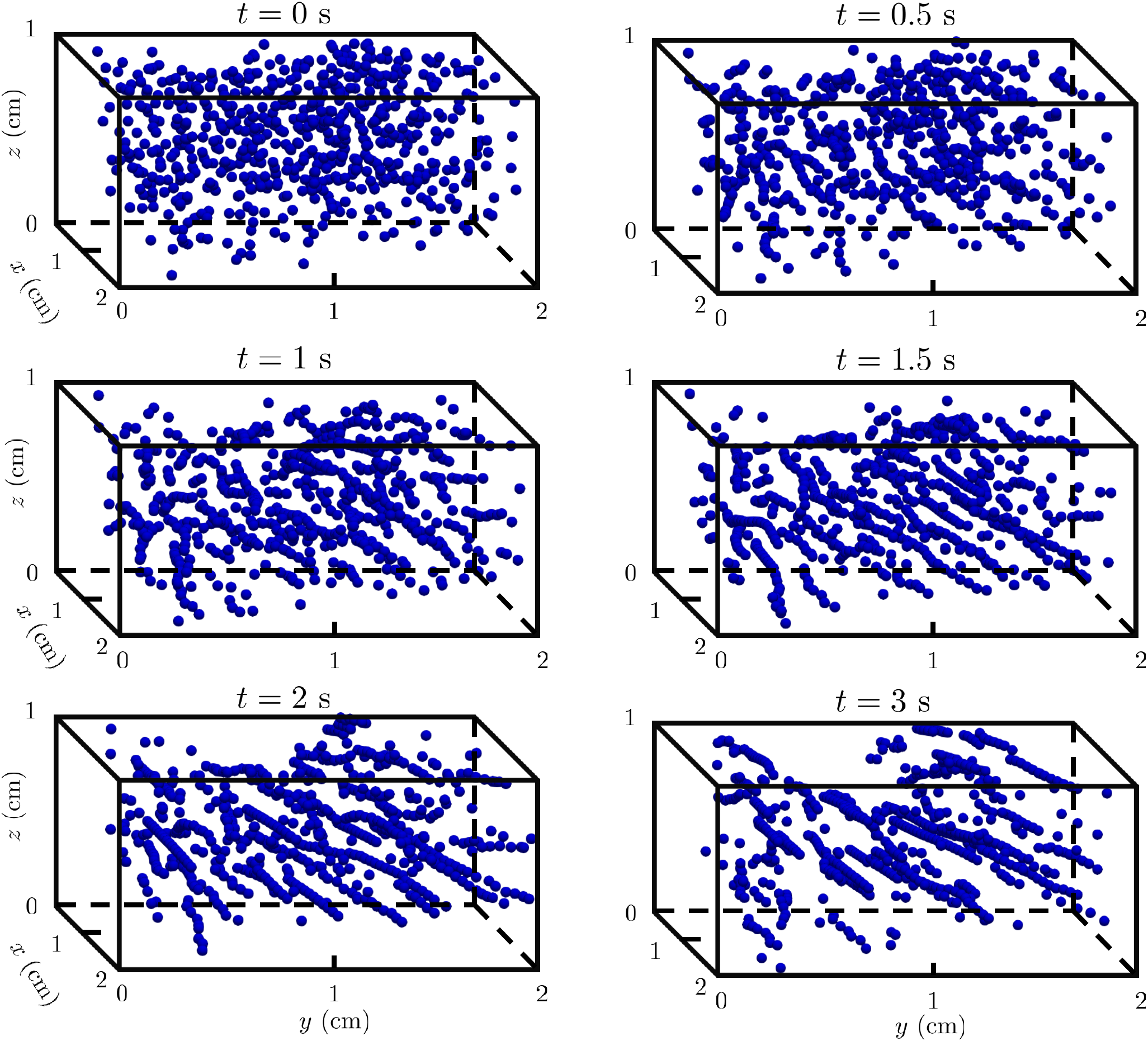}%
\caption[]
{Behavior of a random array of magnetic spheres in a 3D domain with $1.85\%$ particle volume fraction at $t=0,~0.5,~1,~1.5,~2$ and $3$ s under the action of gravity and an external magnetic field applied in the horizontal direction.}
\label{fig:RandomArraysHorizontal}
\end{figure}

Figures~\ref{fig:RandomArraysVertical} and \ref{fig:RandomArraysHorizontal} depict the evolution of the particles at six time instants for the two directions of the imposed external magnetic potential field. It can be seen that with the application of the magnetic field, the
particles become magnetized and acquire a magnetic dipole moment \cite{Han2010}, which promotes the particles to aggregate and form short fragmented chains ($t\leq 1$~s). As time advances, these short chains merge together and form longer chains (i.e., mesoscopic structures) that align in the direction of the applied magnetic field \cite{Han2010}.

\section{Conclusions}
\label{sec:conclusion}

A numerical formulation for fully-resolved simulation of magnetorheological fluids (MRF) consisting of solid magnetic particles suspended in a Newtonian carrier fluid was presented. The implementation was carried out by extending the open-source $CFDEMcoupling$ framework with a force calculation at the particles surface due to the applied external magnetic field, and with the implementation of the fixed and mutual dipole-dipole magnetic models to account for the magnetic interactions between the particles. The overall algorithm procedure solves a second-order differential equation for the magnetic potential field, followed by the flow equations, including the continuity and momentum balance equations, and an immersed boundary algorithm to model the flow around discrete magnetic particles present in the flow domain. This approach guarantees a tight coupling between the dynamics of the fluid and the magnetic solid discrete phase. The coupling is provided by the calculation of the net hydrodynamic and magnetic forces and torques exerted by the fluid on the solid particles. The algorithm subsequently uses the discrete element method to model the particle motion, comprising linear and rotational motions, as well as the particles magnetic moment, which in turn provides new boundary conditions for the fluid domain.

The accuracy and robustness of the proposed FVM-IBM-DEM-MAG algorithm were evaluated using four benchmark studies. First, for the sedimentation of two spheres in a rectangular duct containing a Newtonian fluid without the presence of an external magnetic field (mimicking the drafting-kissing-tumbling, DKT, phenomena), the particles velocity and location were compared with numerical data available in the literature and a good agreement was obtained. The velocity contour profiles of the  particles falling through the Newtonian fluid  distinctly showed several symmetry breaking physical aspects of the non-smooth DKT phenomenon. We also demonstrated the capability of the algorithm to predict the dynamics of two magnetic  particles falling under the action of gravity and an external magnetic field, i.e., the simulation of the DKT benchmark case study but with activating the magnetic forces calculations. For a vertical magnetic field, the particles experience a longer drafting period and form a tight string which does not separate during the rest of the sedimentation process. For a horizontal magnetic field, the particles just tumble as a non-kissing pair and approach the vertical walls of the domain, where the external magnetic field is applied. The FVM-IBM-DEM-MAG solver was also used to study the multi-particle chaining when particles are placed randomly on a 2D-plane. Two area fractions of spheres were tested, $20\%$ and $30\%$, corresponding to 260 and 390 spheres under the effect of gravity and a vertical or horizontal magnetic fields. The snapshots of the particles locations showed that, at the initial instants, short fragmented chains or clusters of particles are formed.  With time advancing, the short chains merge together and form longer column-like chains always aligned with the direction of the externally imposed magnetic field. Finally, the robustness of the FVM-IBM-DEM-MAG solver was tested in a 3D domain, where an array of 580 randomly distributed magnetic particles  were subjected to gravity and a horizontal or vertical magnetic field. Again, the snapshots of the particles location demonstrated the formation of long column-like chains in the direction of the applied magnetic field.
 
In summary, the results presented in this study show that the newly developed code can accurately predict the flow patterns and particle assembly in MRF for a number of benchmark problems.

\section{Acknowledgements}
C. Fernandes acknowledges the support by FEDER funds through the COMPETE 2020 Programme and National Funds through FCT (Portuguese Foundation for Science and Technology) under the projects UID-B/05256/2020 and UID-P/05256/2020.

Salah A. Faroughi would like to acknowledge support by National Science Foundation Partnership for Research and Education in Materials (PREM) (award no. DMR-2122041).

\bibliographystyle{unsrtnat}
\bibliography{references}

\begin{thebibliography}{51}
\providecommand{\natexlab}[1]{#1}
\providecommand{\url}[1]{\texttt{#1}}
\expandafter\ifx\csname urlstyle\endcsname\relax
  \providecommand{\doi}[1]{doi: #1}\else
  \providecommand{\doi}{doi: \begingroup \urlstyle{rm}\Url}\fi

\bibitem[Satoh(2017)]{Akira2017}
A.~Satoh.
\newblock \emph{Modeling of magnetic particle suspensions for simulations}.
\newblock CRC Press Taylor {\&} Francis Group, Florida, 2017.

\bibitem[Kumar et~al.(2019)Kumar, Paul, Raghunathan, and Alex]{Kumar2019}
J.S. Kumar, P.S. Paul, G.~Raghunathan, and D.G. Alex.
\newblock A review of challenges and solutions in the preparation and use of
  magnetorheological fluids.
\newblock \emph{International Journal of Mechanical and Materials Engineering},
  14:\penalty0 13, 2019.
\newblock \doi{https://doi.org/10.1186/s40712-019-0109-2}.

\bibitem[Bullough(1996)]{Bullough1996}
W.A. Bullough.
\newblock \emph{Electro-Rheological Fluids, Magneto-Rheological Suspensions and
  Associated Technology}.
\newblock World Scientific, Singapore, 1996.

\bibitem[Wereley(2013)]{Wereley2013}
N.M. Wereley.
\newblock \emph{Magnetorheology: Advances and Applications}.
\newblock Royal Society of Chemistry, London, 2013.

\bibitem[Kuznetsov et~al.(1999)Kuznetsov, Filippov, Kuznetsov, Gerlivanov,
  Dobrinsky, and Malashin]{Kuz1999}
A.A. Kuznetsov, V.I. Filippov, O.A. Kuznetsov, V.G. Gerlivanov, E.K. Dobrinsky,
  and S.I. Malashin.
\newblock New ferro-carbon adsorbents for magnetically guided transport of
  anti-cancer drugs.
\newblock \emph{Journal of Magnetism and Magnetic Materials}, 194:\penalty0
  22--30, 1999.
\newblock \doi{https://doi.org/10.1016/S0304-8853(98)00568-X}.

\bibitem[Weingart et~al.(2013)Weingart, Vabbilisetty, and Sun]{Wein2013}
J.~Weingart, P.~Vabbilisetty, and X.~Sun.
\newblock Membrane mimetic surface functionalization of nanoparticles: Methods
  and applications.
\newblock \emph{Advances in Colloid and Interface Science}, 197--198:\penalty0
  68--84, 2013.
\newblock \doi{https://doi.org/10.1016/j.cis.2013.04.003}.

\bibitem[Girginova et~al.(2010)Girginova, da~Silva, Lopes, Figueira, Otero,
  Amaral, Pereira, and Trindade]{Girg2010}
P.I. Girginova, A.L.~Daniel da~Silva, C.B. Lopes, P.~Figueira, M.~Otero, V.S.
  Amaral, E.~Pereira, and T.~Trindade.
\newblock Silica coated magnetite particles for magnetic removal of {H}g$^{2+}$
  from water.
\newblock \emph{Journal of Colloid and Interface Science}, 345:\penalty0
  234--240, 2010.
\newblock \doi{https://doi.org/10.1016/j.jcis.2010.01.087}.

\bibitem[Lan et~al.(2013)Lan, Wu, Li, Li, Guo, and Gan]{Lan2013}
S.~Lan, X.~Wu, L.~Li, M.~Li, F.~Guo, and S.~Gan.
\newblock Synthesis and characterization of hyaluronic acid-supported magnetic
  microspheres for copper ions removal.
\newblock \emph{Colloids and Surfaces A: Physicochemical and Engineering
  Aspects}, 425:\penalty0 42--50, 2013.
\newblock \doi{https://doi.org/10.1016/j.colsurfa.2013.02.059}.

\bibitem[Chow(2006)]{Chow2006}
T.L. Chow.
\newblock \emph{Introduction to electromagnetic theory: a modern perspective}.
\newblock Jones {\&} Bartlett Learning, Massachusetts, 2006.

\bibitem[Grant and Phillips(2008)]{Grant2008}
I.S. Grant and W.R. Phillips.
\newblock \emph{Electromagnetism}.
\newblock John Wiley {\&} Sons, New York, 2008.

\bibitem[Kang et~al.(2008)Kang, Hulsen, den Toonder, Anderson, and
  Meijer]{Kang2008}
T.G. Kang, M.A. Hulsen, J.M.J. den Toonder, P.D. Anderson, and H.E.H. Meijer.
\newblock A direct simulation method for flows with suspended paramagnetic
  particles.
\newblock \emph{Journal of Computational Physics}, 227:\penalty0 4441--4458,
  2008.
\newblock \doi{https://doi.org/10.1016/j.jcp.2008.01.005}.

\bibitem[Hayes et~al.(2001)Hayes, Polson, and Garcia]{Hayes2001}
M.A. Hayes, N.P. Polson, and A.A. Garcia.
\newblock Active control of dynamic supraparticle structures in microchannels.
\newblock \emph{Langmuir}, 17:\penalty0 2866--2871, 2001.
\newblock \doi{https://doi.org/10.1021/la001655g}.

\bibitem[Melle and Martin(2003)]{Melle2003}
S.~Melle and J.E. Martin.
\newblock Chain model of a magnetorheological suspension in a rotating field.
\newblock \emph{Journal of Chemical Physics}, 118:\penalty0 9875--9881, 2003.
\newblock \doi{https://doi.org/10.1063/1.1570817}.

\bibitem[Keaveny and Maxey(2008)]{Keaveny2008}
E.E. Keaveny and M.R. Maxey.
\newblock Modeling the magnetic interactions between paramagnetic beads in
  magnetorheological fluids.
\newblock \emph{Journal of Computational Physics}, 227:\penalty0 9554--9571,
  2008.
\newblock \doi{https://doi.org/10.1016/j.jcp.2008.07.008}.

\bibitem[Han et~al.(2010)Han, Feng, and Owen]{Han2010}
K.~Han, Y.T. Feng, and D.R.J. Owen.
\newblock Three-dimensional modelling and simulation of magnetorheological
  fluids.
\newblock \emph{International Journal for Numerical Methods in Engineering},
  84:\penalty0 1273--1302, 2010.
\newblock \doi{https://doi.org/10.1002/nme.2940}.

\bibitem[Stokes(1901)]{Stokes1851}
G.G. Stokes.
\newblock On the effect of internal friction of fluids on the motion of
  pendulums.
\newblock In G.G. Stokes, editor, \emph{Mathematical and Physical Papers}.
  Cambridge University Press, Cambridge, 1901.
\newblock \doi{https://doi.org/10.1017/CBO9780511702266.002}.

\bibitem[Ke et~al.(2017)Ke, Shu, Zhang, and Yuan]{Ke2017}
C.-H. Ke, S.~Shu, H.~Zhang, and H.-Z. Yuan.
\newblock {LBM}-{IBM}-{DEM} modelling of magnetic particles in a fluid.
\newblock \emph{Powder Technology}, 314:\penalty0 264--280, 2017.
\newblock \doi{http://doi.org/10.1016/j.powtec.2016.08.008}.

\bibitem[Zhang et~al.(2019)Zhang, Zhou, and Shao]{Zhang2019}
S.~Zhang, J.~Zhou, and C.~Shao.
\newblock Numerical investigation on yielding phenomena of magnetorheological
  fluid flowing through microchannel governed by transverse magnetic field.
\newblock \emph{Physics of Fluids}, 31:\penalty0 022005, 2019.
\newblock \doi{https://doi.org/10.1063/1.5079624}.

\bibitem[Zhou et~al.(2019)Zhou, Zhang, Tian, and Shao]{Zhou2019}
J.-Feng Zhou, S.~Zhang, F.~Tian, and C.-Lei Shao.
\newblock Simulation of oscillation of magnetic particles in 3{D} microchannel
  flow subjected to alternating gradient magnetic field.
\newblock \emph{Journal of Magnetism and Magnetic Materials}, 473:\penalty0
  32--41, 2019.
\newblock \doi{https://doi.org/10.1016/j.jmmm.2018.10.028}.

\bibitem[Leps and Hartzell(2021)]{Leps2021}
T.~Leps and C.~Hartzell.
\newblock High fidelity, discrete element method simulation of
  magnetorheological fluids using accurate particle size distributions in
  {LIGGGHTS} extended with mutual dipole method.
\newblock \emph{Materials Research Express}, 8:\penalty0 085701, 2021.
\newblock \doi{https://doi.org/10.1088/2053-1591/ac113c}.

\bibitem[{C}omputing {G}mb{H}(2015)]{GMBH2015Contact}
{DCS} {C}omputing {G}mb{H}.
\newblock {LIGGGHTS} public documentation, 2015.
\newblock URL \url{https://www.cfdem.com/media/DEM/docu/gran_model_hertz.html}.

\bibitem[Tajfirooz et~al.(2021)Tajfirooz, Meijer, Dellaert, Meulenbroek,
  Zeegers, and Kuerten]{Tajfirooz2021}
S.~Tajfirooz, J.G. Meijer, R.A. Dellaert, A.M. Meulenbroek, J.C.H. Zeegers, and
  J.G.M. Kuerten.
\newblock Direct numerical simulation of magneto-{A}rchimedes separation of
  spherical particles.
\newblock \emph{Journal of Fluid Mechanics}, 910:\penalty0 A52, 2021.
\newblock \doi{https://doi.org/10.1017/jfm.2020.1001}.

\bibitem[Zhou et~al.(2010)Zhou, Kuang, Chu, and Yu]{Zhou2010}
Z.~Zhou, S.~Kuang, K.~Chu, and A.~Yu.
\newblock Discrete particle simulation of particle-fluid flow: Model
  formulations and their applicability.
\newblock \emph{Journal of Fluid Mechanics}, 661:\penalty0 482--510, 2010.
\newblock \doi{https://doi.org/10.1017/S002211201000306X}.

\bibitem[Fernandes et~al.(2018)Fernandes, Semyonov, Ferrás, and
  Nóbrega]{Fernandes2018}
C.~Fernandes, D.~Semyonov, L.L. Ferrás, and J.M. Nóbrega.
\newblock Validation of the {CFD}--{DPM} solver {DPMF}oam in {O}pen{FOAM}
  through analytical, numerical and experimental comparisons.
\newblock \emph{Granular Matter}, 20:\penalty0 64, 2018.
\newblock \doi{https://doi.org/10.1007/s10035-018-0834-x}.

\bibitem[Osher and Fedkiw(2003)]{Osher2003}
S.~Osher and S.~Fedkiw.
\newblock \emph{Level Set Methods and Dynamic Implicit Surfaces}.
\newblock Springer-Verlag, New York, 2003.

\bibitem[Jackson(1999)]{Jackson1999}
J.D. Jackson.
\newblock \emph{Classical Electrodynamics}.
\newblock John Wiley {\&} Sons, New York, 1999.

\bibitem[Gontijo and Cunha(2017)]{gontijo2017numerical}
R.G. Gontijo and F.R. Cunha.
\newblock Numerical simulations of magnetic suspensions with hydrodynamic and
  dipole-dipole magnetic interactions.
\newblock \emph{Physics of Fluids}, 29\penalty0 (6):\penalty0 062004, 2017.
\newblock \doi{https://doi.org/10.1063/1.4986083}.

\bibitem[Ferraro(1961)]{Ferraro1961}
V.C.A. Ferraro.
\newblock \emph{Electromagnetic Theory}.
\newblock The Athlone Press, London, 1961.

\bibitem[{L}arge-scale {A}tomic/{M}olecular {M}assively~{P}arallel
  {S}imulator(2022)]{lammpsDipole}
{L}arge-scale {A}tomic/{M}olecular {M}assively~{P}arallel {S}imulator.
\newblock Dipole-dipole, 2022.
\newblock URL \url{https://docs.lammps.org/pair_dipole.html}.

\bibitem[Glowinski et~al.(2001)Glowinski, Pan, Hesla, Joseph, and
  Périaux]{Glowinsky2000}
R.~Glowinski, T.W. Pan, T.I. Hesla, D.D. Joseph, and J.~Périaux.
\newblock A fictitious domain approach to the direct numerical simulation of
  incompressible viscous flow past moving rigid bodies: Application to
  particulate flow.
\newblock \emph{Journal of Computational Physics}, 169:\penalty0 363--426,
  2001.
\newblock \doi{https://doi.org/10.1006/jcph.2000.6542}.

\bibitem[Fernandes et~al.(2019)Fernandes, Faroughi, Carneiro, Nóbrega, and
  McKinley]{Fernandes2019}
C.~Fernandes, S.A. Faroughi, O.S. Carneiro, J.M. Nóbrega, and G.H. McKinley.
\newblock Fully-resolved simulations of particle-laden viscoelastic fluids
  using an immersed boundary method.
\newblock \emph{Journal of Non-Newtonian Fluid Mechanics}, 266:\penalty0
  80--94, 2019.
\newblock \doi{https://doi.org/10.1016/j.jnnfm.2019.02.007}.

\bibitem[Renzo and Maio(2004)]{Renzo2004}
A.~Di Renzo and F.P.~Di Maio.
\newblock Comparison of contact-force models for the simulation of collisions
  in {DEM}-based granular flow codes.
\newblock \emph{Chemical Engineering Science}, 59:\penalty0 525--541, 2004.
\newblock \doi{https://doi.org/10.1016/j.ces.2003.09.037}.

\bibitem[Kloss et~al.(2012)Kloss, Goniva, Hager, Amberger, and
  Pirker]{Kloss2012}
C.~Kloss, C.~Goniva, A.~Hager, S.~Amberger, and S.~Pirker.
\newblock Models, algorithms and validation for opensource {DEM} and
  {CFD}-{DEM}.
\newblock \emph{Progress in Computational Fluid Dynamics}, 12:\penalty0
  140--152, 2012.
\newblock \doi{https://doi.org/10.1504/PCFD.2012.047457}.

\bibitem[Cundall and Strack(1979)]{Cundall197947}
P.A. Cundall and O.D.L. Strack.
\newblock A discrete numerical model for granular assemblies.
\newblock \emph{G\'eotechnique}, 29:\penalty0 47--65, 1979.
\newblock \doi{https://doi.org/10.1680/geot.1979.29.1.47}.

\bibitem[Nezami et~al.(2004)Nezami, Hashash, Zhao, and
  Ghaboussi]{nezami2004fast}
E.G. Nezami, Y.M.A. Hashash, D.~Zhao, and J.~Ghaboussi.
\newblock A fast contact detection algorithm for 3-{D} discrete element method.
\newblock \emph{Computers and Geotechnics}, 31\penalty0 (7):\penalty0 575--587,
  2004.
\newblock \doi{https://doi.org/10.1016/j.compgeo.2004.08.002}.

\bibitem[Lu et~al.(2021)Lu, Gao, Dietiker, Shahnam, and Rogers]{lu2021machine}
L.~Lu, X.~Gao, J.-F. Dietiker, M.~Shahnam, and W.A. Rogers.
\newblock Machine learning accelerated discrete element modeling of granular
  flows.
\newblock \emph{Chemical Engineering Science}, 245:\penalty0 116832, 2021.
\newblock \doi{https://doi.org/10.1016/j.ces.2021.116832}.

\bibitem[Hager et~al.(2014)Hager, Kloss, Pirker, and Goniva]{Hager2014}
A.~Hager, C.~Kloss, S.~Pirker, and C.~Goniva.
\newblock Parallel resolved open source {CFD}-{DEM}: method, validation and
  application.
\newblock \emph{The Journal of Computational Multiphase Flows}, 6:\penalty0
  13--27, 2014.
\newblock \doi{https://doi.org/10.1260/1757-482X.6.1.13}.

\bibitem[Aycock et~al.(2017)Aycock, Campbell, Manning, and Craven]{Kenneth2017}
K.I. Aycock, R.L. Campbell, K.B. Manning, and B.A. Craven.
\newblock A resolved two-way coupled {CFD}/6-{DOF} approach for predicting
  embolus transport and the embolus-trapping efficiency of {IVC} filters.
\newblock \emph{Biomechanics and Modeling in Mechanobiology}, 16:\penalty0
  851--869, 2017.
\newblock \doi{https://doi.org/10.1007/s10237-016-0857-3}.

\bibitem[Verlet(1967)]{Verlet1967}
L.~Verlet.
\newblock Computer experiments on classical fluids {I}. {T}hermodynamical
  properties of {L}ennard-{J}ones molecules.
\newblock \emph{Physical Review}, 159:\penalty0 98--103, 1967.
\newblock \doi{https://doi.org/10.1103/PhysRev.159.98}.

\bibitem[Issa(1986)]{Issa1986}
R.I. Issa.
\newblock Solution of the implicitly discretized fluid flow equations by
  operator-splitting.
\newblock \emph{Journal of Computational Physics}, 62:\penalty0 40--65, 1986.
\newblock \doi{https://doi.org/10.1016/0021-9991(86)90099-9}.

\bibitem[{CFDEM}coupling(2011)]{CFDDEMcoupling2011}
{CFDEM}coupling.
\newblock {CFDEM} project, 2011.
\newblock URL \url{https://www.cfdem.com/cfdemcoupling}.

\bibitem[Fortes et~al.(1987)Fortes, Joseph, and Lundgren]{Fortes1987}
F.~Fortes, D.D. Joseph, and T.S. Lundgren.
\newblock Non-linear mechanics of fluidization of beds of spherical particles.
\newblock \emph{Journal of Fluid Mechanics}, 177:\penalty0 467--483, 1987.
\newblock \doi{https://doi.org/10.1017/S0022112087001046}.

\bibitem[Hu et~al.(1992)Hu, Joseph, and Crochet]{Hu1992}
H.H. Hu, D.D. Joseph, and M.J. Crochet.
\newblock Direct simulation of fluid particle motions.
\newblock \emph{Theoretical and Computational Fluid Dynamics}, 3:\penalty0
  285--306, 1992.
\newblock \doi{https://doi.org/10.1007/BF00717645}.

\bibitem[Johnson and Tezduyar(1996)]{Johnson1992}
A.A. Johnson and T.E. Tezduyar.
\newblock Simulation of multiple spheres falling in a liquid-filled tube.
\newblock \emph{Computer Methods in Applied Mechanics and Engineering},
  134:\penalty0 351--373, 1996.
\newblock \doi{https://doi.org/10.1016/0045-7825(95)00988-4}.

\bibitem[Feng et~al.(1994)Feng, Hu, and Joseph]{Feng1994}
J.~Feng, H.H. Hu, and D.D. Joseph.
\newblock Direct simulation of initial value problems for the motion of solid
  bodies in a {N}ewtonian fluid. {P}art 1. {S}edimentation.
\newblock \emph{Journal of Fluid Mechanics}, 261:\penalty0 95--134, 1994.
\newblock \doi{https://doi.org/10.1017/S0022112094000285}.

\bibitem[Jasak(2009)]{Jasak2009}
H.~Jasak.
\newblock Dynamic mesh handling in {O}pen{FOAM}.
\newblock In \emph{47th {AIAA} {A}erospace {S}ciences {M}eeting}, Orlando,
  Florida, 2009.

\bibitem[Huang et~al.(1994)Huang, Feng, and Joseph]{Huang1994}
P.Y. Huang, J.~Feng, and D.D. Joseph.
\newblock The turning couples on an elliptic particle settling in a vertical
  channel.
\newblock \emph{Journal of Fluid Mechanics}, 271:\penalty0 1--16, 1994.
\newblock \doi{https://doi.org/10.1017/S0022112094001667}.

\bibitem[Ritz and Caltagirone(1999)]{Ritz1999}
J.B. Ritz and J.P. Caltagirone.
\newblock A numerical continuous model for the hydrodynamics of fluid particle
  systems.
\newblock \emph{International Journal for Numerical Methods in Fluids},
  30:\penalty0 1067--1090, 1999.
\newblock
  \doi{https://doi.org/10.1002/(SICI)1097-0363(19990830)30:8<1067::AID-FLD881>3.0.CO;2-6}.

\bibitem[Ly et~al.(1999)Ly, Reitich, Jolly, Banks, and Ito]{Ly1999}
H.V. Ly, F.~Reitich, M.R. Jolly, H.T. Banks, and K.~Ito.
\newblock Simulations of particle dynamics in magnetorheological fluids.
\newblock \emph{Journal of Computational Physics}, 155:\penalty0 160--177,
  1999.
\newblock \doi{https://doi.org/10.1006/jcph.1999.6335}.

\bibitem[Mohebi et~al.(1996)Mohebi, Jamasbi, and Liu]{Mohebi1996}
M.~Mohebi, N.~Jamasbi, and J.~Liu.
\newblock Simulation of the formation of nonequilibrium structures in
  magnetorheological fluids subject to an external magnetic field.
\newblock \emph{Physical Review E}, 54:\penalty0 5407--5413, 1996.
\newblock \doi{https://doi.org/10.1103/PhysRevE.54.5407}.

\bibitem[Fermigier and Gast(1992)]{Fermigier1992}
M.~Fermigier and A.P. Gast.
\newblock Structure evolution in a paramagnetic latex suspension.
\newblock \emph{Journal of Colloid Interface Sciences}, 154:\penalty0 522--539,
  1992.
\newblock \doi{https://doi.org/10.1016/0021-9797(92)90165-I}.

\end{thebibliography}
\end{document}